# Three-dimensional nanoscale control of magnetism in crystalline Yttrium Iron Garnet

Valerio Levati[1†], Matteo Vitali[1†], Andrea Del Giacco[1], Nicola Pellizzi[1], Raffaele Silvani[2], Luca Ciaccarini Mavilla[2], Marco Madami[2], Irene Biancardi[1], Davide Girardi[1], Matteo Panzeri[1], Piero Florio[1], Maria Cocconcelli[1], David Breitbach[3], Philipp Pirro[3], Ludovica Rovatti[4], Nora Lecis[4], Federico Maspero[1], Riccardo Bertacco[1], Giacomo Corrielli[5], Roberto Osellame[5], Valeria Russo[6], Andrea Li Bassi[6], Silvia Tacchi[7*], Daniela Petti[1*], Edoardo Albisetti[1*]

[1] Dipartimento di Fisica, Politecnico di Milano, Via Giuseppe Colombo, 81 Milano 20133, Italy.
[2] Dipartimento di Fisica e Geologia, Università di Perugia, Via A. Pascoli, Perugia I-06123, Italy.
[3] Fachbereich Physik and Landesforschungszentrum OPTIMAS, Rheinland-Pfälzische Technische Universität Kaiserslautern-Landau, Kaiserslautern, Germany.
[4] Dipartimento di Ingegneria Meccanica, Politecnico di Milano, Milan 20156, Italy.
[5] Istituto di Fotonica e Nanotecnologie - Consiglio Nazionale delle Ricerche (IFN-CNR), piazza Leonardo da Vinci 32, 20133 Milano, Italy.
[6] Dipartimento di Energia, Laboratorio Materiali Micro e Nanostrutturati, Politecnico di Milano, via Ponzio 34/3, I-20133 Milano, Italy.
[7] Istituto Officina dei Materiali del CNR (CNR-IOM), Sede Secondaria di Perugia c/o Dipartimento di Fisica e Geologia, Università di Perugia, Perugia I-06123, Italy.

[†]These authors contributed equally: Valerio Levati and Matteo Vitali

*Corresponding author.
Email: silvia.tacchi@iom.cnr.it (ST); daniela.petti@polimi.it (DP); edoardo.albisetti@polimi.it (EA)

**The exceptional magnetic, optical and phononic properties of Yttrium Iron Garnet (YIG) make it unique for spin-wave based and photonic applications. Yet, nanostructuring crystalline YIG and manipulating its magnetism in a non-destructive way is an outstanding challenge, and so far mostly limited to two-dimensional capabilities.**

**Here, we show that irradiation of single-crystal YIG films with a focused UV laser drives a stable, giant enhancement of the perpendicular magnetic anisotropy, preserving the crystalline quality. This modulation is highly confined at the nanoscale in both the lateral and vertical directions, and its extension within the volume can be finely tuned with a continuous depth-control.**

**By harnessing these three-dimensional anisotropy profiles, we demonstrate a large tuning of the spin-wave band structure, volume spatial localization, and non-reciprocity, realizing proof-of-principle 3D magnonic crystals.**

**This straightforward, single-step, laser nanofabrication of three-dimensional magnetic systems based on crystalline YIG thin films opens the way to design novel functions in magnonic and magneto-optic devices.**

## INTRODUCTION

Ferrimagnetic insulating garnets such as crystalline Yttrium Iron Garnet (YIG, $Y_3Fe_5O_{12}$) are key across spintronics to photonics and phononics, thanks to their exceptional magnetic, magneto-optical and mechanical properties. The record-low magnetic damping of crystalline YIG, for example, allows the propagation of spin waves across millimetre distances, making it the material of choice for energy-efficient signal transmission and computing in the field of magnonics[1–3], as well as for the study of fundamental physical phenomena such as Bose-Einstein condensation[4], and quantum computing in hybrid systems[5]. In this framework, the realization of three-dimensional magnetic nanostructures[6,7], such as metamaterials[8], curvilinear and interconnected nanoarchitectures[9–11], chiral and topological spin textures[12–14], enables the discovery of novel effects and the design of new functions in nanodevices. Prototypical examples in the field of spintronics include 3D domain-wall memories[15,16] and logic[17], laying the foundations for three-dimensional integrated magnetic platforms. In magnonics[18,19], the three-dimensional control of magnetism unlocks new possibilities for engineering non-reciprocal phenomena in magnonic circuits[20–22], or to exploit the spin-wave band gaps in 3D magnonic crystals[23,24]. Moreover, exploiting the third dimension allows for dense scalable networks[25–27], vertically coupled systems[28–30] and both spatial and frequency multiplexing for signal processing[31,32]. Eventually, this additional degree of freedom can favour emerging coupling phenomena in hybrid and quantum systems[33–35].

For all these applications, the nanoscale manipulation of the magnetic properties of YIG and the nanofabrication of crystalline YIG-based devices is essential. Yet, high-quality YIG nanopatterning presents formidable challenges, and methodologies natively capable of three-dimensional patterning are missing. This is due to the complex crystalline structure of YIG, which often leads to a degradation of the material properties following conventional 2D nanofabrication[36,37] via etching[38,39], lift-off[40], or irradiation[41–43], and is not compatible with native direct-write three-dimensional nanofabrication techniques such as Focused Electron or Ion Beam Induced Deposition (FEBID/FIBID)[44–46], and Two-Photon Lithography (TPL)[47]. As a result, so far, the realization of three-dimensional YIG structures is limited to specific geometries such as suspended resonators[48], or meandering structures[49] realized using complex lithographic processes. The direct writing of arbitrary-shaped three-dimensional magnetic nanopatterns in crystalline YIG is thus an open challenge.

Here, we exploit a sharp permanent transition in the magnetic properties induced by a focused continuous-wave UV laser, for creating three-dimensional modulations of the magnetic properties of single-crystal YIG(111) thin films, without altering significantly the film topography and crystalline structure. In particular, the perpendicular magnetic anisotropy (PMA) of YIG sharply rises to $\sim 30$ times its pristine value in a confined volume down to 100 nm in size, whose depth can be continuously tuned with the laser power, allowing to induce complex three-dimensional anisotropy profiles and spin textures throughout the volume of the film. Furthermore, by irradiating through the transparent GGG substrate, we realize buried patterns localized at the bottom YIG/GGG surface, and align them with the top patterns in a cross-bar array configuration. With the support of structural characterisation techniques and finite-element numerical simulation, we developed a model for the interaction of the laser light with the YIG film. The induced three-dimensional patterns are explained in terms of depth-dependent local heating due to linear laser absorption, and subsequent re-crystallization in a strained state characterized by enhanced perpendicular magnetic anisotropy. By studying the magnetization dynamics in the 3D nanostructured system, we observe large modulations of the spin-wave band structure and the emergence of novel non-reciprocal spin-wave modes, whose dispersion and spatial localization are tunable with the laser power. Finally, we harness this capability to realize direct-write three-dimensional magnonic crystals.

## RESULTS

### Three-dimensional magnetic nanopatterning

The studied system consists of 1 μm-thick single-crystal YIG films grown via liquid phase epitaxy on a GGG(111) substrate (See Supplementary Fig. 8). Fig. 1a illustrates the irradiation process, where a focused UV 405 nm continuous-wave laser is raster scanned in ambient conditions on the sample surface[50] (see Methods).

We first irradiated 5 μm-size dot-shaped regions with different laser power, and measured the magnetic domain structure at remanence via Magnetic Force Microscopy (MFM) (Fig. 1c). Pristine YIG exhibits weak stripe domains with a periodicity of ∼ 1.6 μm, consisting of an alternating small out-of-plane component of the magnetization, while the magnetization predominantly lies in the sample plane. Conversely, upon irradiation, we identified two regimes, separated by a sharp threshold laser power $P_{\text{th}} \sim 27.5$ mW. Below threshold, no major change is observed, apart from a slight reduction of the MFM magnetic contrast. Conversely, above threshold, we observed the onset of narrow stripe domains with periodicity down to ∼ 0.6 μm, which then gradually increases with the laser power. Importantly, such a threshold leads to extremely sharp pattern edges where the magnetic properties change abruptly, and to the possibility of realizing nanosized patterns well below the nominal laser spot size of 1.2 μm, by irradiating around the threshold power. For example, Fig. 1b shows mono-domain dot arrays down to ~ 100 nm in size, realized by irradiating with single 50 μs-long pulses. By controlling the pattern geometry and irradiation power, a rich variety of magnetic features such as circular, bubble domains or zig-zag stripes, can be stabilized at remanence and manipulated with external fields (Supplementary Fig. 3). For all these patterns, we set a fixed laser irradiation time to ensure precise control of the magnetic properties while preserving the surface quality of the exposed area. The effect of laser irradiation time is further discussed in Supplementary Fig. 4.

So far, a uniform irradiation was employed within each pattern. However, by varying the laser power point-by-point in a "grayscale" fashion, we obtained a graded magnetic nanomaterial. Fig. 1d shows the MFM domain structure (topography in Supplementary Fig. 5) obtained with a sinusoidal spatial variation of the laser power, as shown in the corresponding spatial power map of Fig. 1e. Such spatially varying magnetic properties give rise to a complex magnetic landscape, where large stripe domains coexist with narrow stripes at low power, then gradually disappear as the power increases (dashed rectangle from left to right). As explained in depth below, such complex texture originates from the three-dimensional modulation of the thickness of the modified magnetic volume, induced by the spatially-varying laser power (schematically shown in Fig. 1f).

Finally, we demonstrate the direct realization of buried structures by irradiating through the transparent GGG substrate. Fig. 1g shows the MFM image of a 2 μm-wide cross-bar array, where top vertical tracks are aligned with horizontal buried tracks localized at the GGG/YIG interface, as schematically shown in Fig. 1h. Combining three-dimensional tuning with straightforward access to the buried bottom interface enables the realization of vertically coupled devices and complex 3D networks in thin YIG films. Besides, irradiation and alignment through the substrate also enables to access YIG films where metallic functional layers or nanostructures are already present on the surface.

In addition to enabling flip-chip patterning, the GGG substrate plays a crucial role in ensuring the high magnetic quality of the YIG film, thanks to its low lattice mismatch, which results in a unstrained pristine state. However, the substrate does not directly influence the patterning mechanism, as the optical power is fully absorbed within the YIG layer (see Supplementary Fig. 2). Its role may become more significant in the case of thinner YIG films, where heat dissipation and thermal gradients are more strongly affected by the substrate's thermal properties.

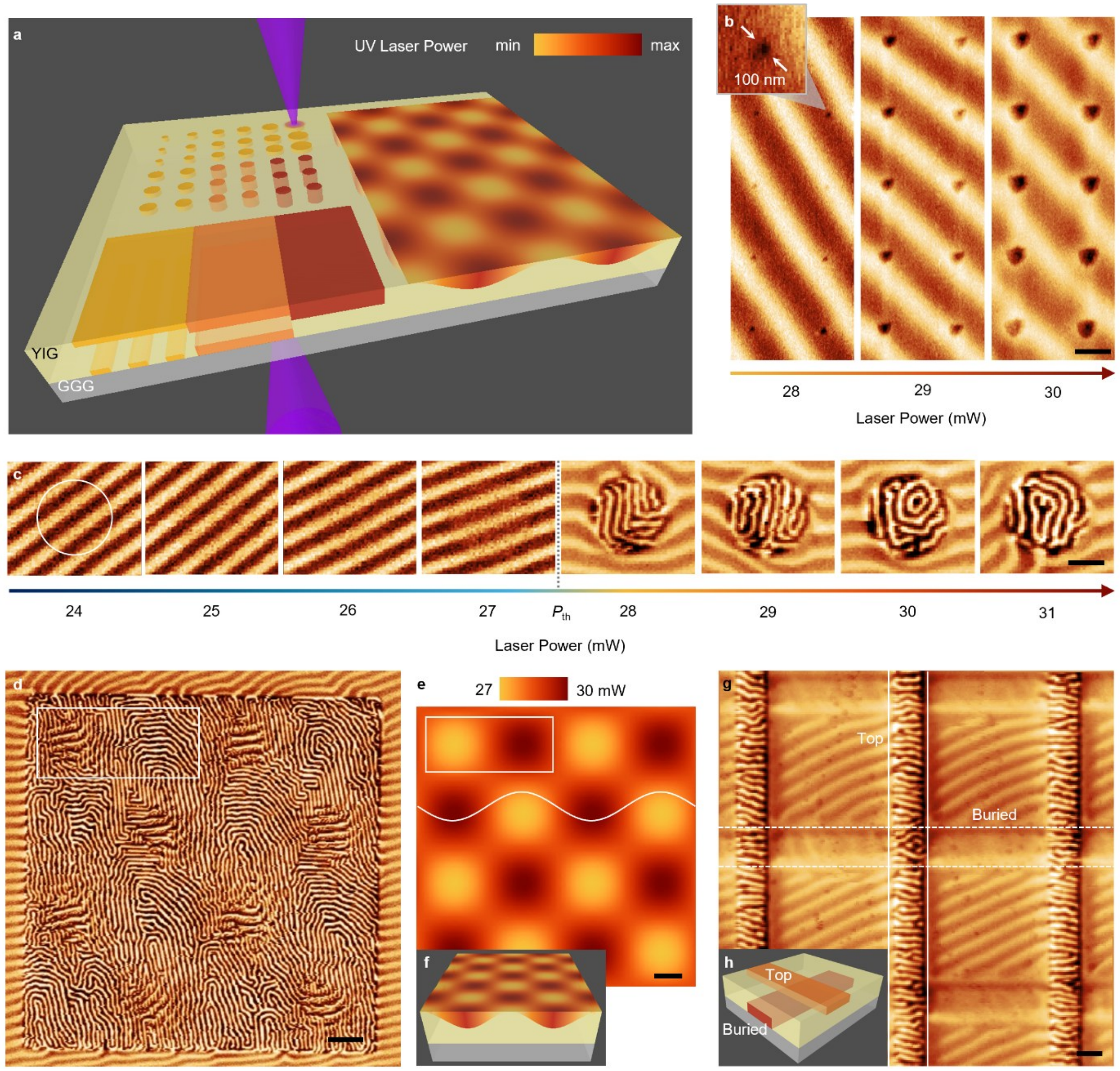

**Figure 1. Three-dimensional control of magnetism in single-crystal YIG films via laser irradiation. a,** Sketch of the patterning process. A focused continuous-wave UV laser is scanned on a single-crystal YIG film in ambient conditions, leading to three-dimensional magnetic patterns with tunable magnetic properties. **b,** Magnetic domain structure of arrays of nanodots down to 100 nm in lateral size, obtained via single-shot irradiation with different laser power, measured via Magnetic Force Microscopy (MFM) at remanence. Scale bar: 1 μm. **c,** 5 μm-size circular areas obtained by raster-scanning at different laser power. Above the threshold power $P_{th}$ complex patterns of narrow stripe domains are stabilized, featuring a spatially sharp reduction in periodicity, tunable with the laser power, and enhanced MFM magnetic contrast. Scale bar: 2 μm. **d-f,** Continuous three-dimensional patterning, obtained irradiating with a spatially varying sinusoidal power profile coded in the power map (**e**), resulting in a continuous point-by-point depth-control (as sketched in **f**). The MFM (**d**, white rectangle) shows the gradual spatial variation of the domain structure enabled by the 3D modulation of the magnetic properties. Scale bars: 5 μm. **g, h,** Cross-bar array consisting of 2 μm-wide tracks localized at the top and bottom surfaces of the YIG film. The buried tracks were obtained by irradiating through the transparent GGG substrate. Scale bar: 2 μm.

**Magnetic properties and static micromagnetic modelling**

To better understand and quantify our observations, we acquired local hysteresis loops of the irradiated areas via Magneto-Optical Kerr Effect (MOKE). Fig. 2a, b report the in-plane loops as a function of the patterning laser power. As expected for weak stripe domains[51], the pristine loop (black line) has a switching-type behaviour at low fields, and a gradual saturation up to $\mu_0 H_{sat} = 3.3$ mT, associated to the reorientation of the out-of-plane component of the stripes. Strikingly, we observe that $H_{sat}$ increases with the laser power up to 67 mT (Fig. 2c), i.e. more than 20-fold with respect to the pristine area indicating an increase of the out-of-plane anisotropy. Conversely, the MOKE signal amplitude does not change appreciably, suggesting that the saturation magnetization does not change significantly (see Supplementary Fig. 6). Besides, the MOKE out-of-plane hysteresis loops, shown in Supplementary Fig. 7, are consistent with an increase of PMA and are in good agreement with the corresponding micromagnetic simulations. In order to correlate such macroscopic magnetic properties with the domain nanostructure, we acquired MFM images as a function of the external in-plane field (insets of Fig. 2a). The irradiated areas (red squares) feature narrow stripe domains which are still visible at a 50 mT field, consistently with a large increase of the perpendicular magnetic anisotropy upon irradiation.

In order to assess the enhancement in perpendicular magnetic anisotropy, we modelled our system via micromagnetic simulations[52] (Fig. 2d-f). For the pristine YIG film, a good agreement is obtained setting a saturation magnetization $M_S = 140$ kA/m" and an exchange stiffness $A = 4$ pJ/m, typical of YIG, and the perpendicular uniaxial anisotropy energy density to $K_U = 200$ J/m$^3$. Based on the SEM cross-sections of Fig. 3a (see section below), the irradiated samples were simulated considering a bicomponent system where the bottom volume is pristine YIG, while the top ”modified” region has an enhanced value of $K_U$. As shown in the simulated spin texture of Fig. 2e-f, the onset of narrow stripe domains in the top layer is well reproduced setting $K_U = 6000$ J/m$^3$ in the top volume, i.e. about 30 times the pristine value. Furthermore, both the increase of the stripes periodicity[53] and the gradual disappearance of the bottom stripe domains with the laser power (as shown in Fig. 1f) are nicely explained by the gradual increase of the thickness of the modified volume, in agreement with the experimental SEM characterisation.

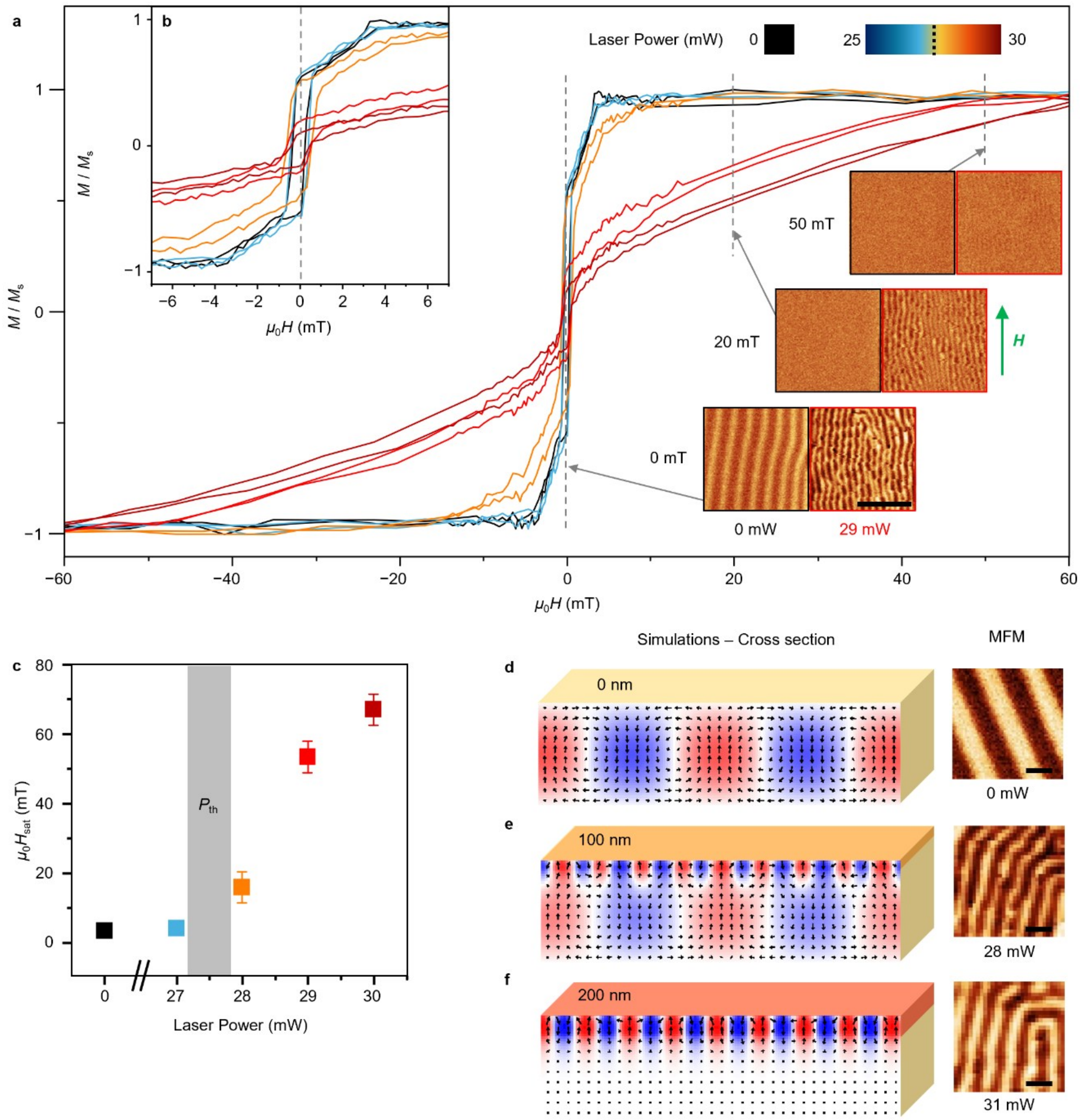

**Figure 2. Giant tuning of perpendicular magnetic anisotropy and domain structure. a, b,** Local in-plane hysteresis loop of areas irradiated with different laser power, measured by Magneto-Optical Kerr Effect (**a**), and zoomed-in view of the central region of the loops (**b**). Below $P_{th}$ (blue curve) we observe only minor changes with respect to the pristine film (black curve). Above $P_{th}$ (orange, red curves) instead the loops are elongated, with enhanced saturation field. The insets of panel **a** show the magnetic domain structure of the pristine (black border) vs irradiated (red border) areas, at different in-plane external fields, corresponding to the dashed lines in the hysteresis loop. The stripe domains disappear below 5 mT in the pristine region and above 50 mT in the irradiated region. Scale bar: 5 μm. **c.** Measured saturation field $H_{sat}$ as a function of the irradiation power, showing a ~ 20-fold increase with respect to pristine areas. **d-f,** Micromagnetic simulations of the three-dimensional spin texture, and corresponding MFM images, showing the transition from weak large stripe domains in the pristine film (**d**), to narrow stripes localized within the top irradiated surface (**f**). Scale bars: 1 μm.

## Structural and optical characterisation

To directly investigate the properties of the patterns and their three-dimensionality, we performed an in-depth structural and optical characterisation. Fig. 3a shows a Scanning Electronic Microscope (SEM) 20 deg-tilted cross-section acquired on a sample irradiated at different powers and subsequently cleaved across the patterns, where the YIG sidewall and top surface are visible. The images reveal a sharp contrast between the brighter smooth sidewall of the bottom pristine YIG and the top modified volume, whose depth can be continuously varied by irradiation. The corresponding relation between laser power and modified depth is presented in Fig. 3b, with a roughly linear trend in the studied power range, in good agreement with numerical simulations of Supplementary Fig. 2. Interestingly, while the patterns' surface appears smooth, the top modified volume features corrugated sidewalls, consistently with the effect of residual strain relaxation upon cutting, suggesting the presence of strain within the modified volume. Indeed, for laser power larger than 30 mW, the patterns initially appear mechanically stable, but can develop cracks due to strain relaxation within minutes after irradiation.

Additionally, to better investigate the structural properties, we acquired micro-Raman spectra (Fig. 3e) on regions irradiated with different laser power. The spectra are remarkably similar to that of the pristine YIG, indicating no major structural changes in the whole power range, and that high crystalline quality and lattice orientation is preserved (see also Electron Backscatter Diffraction data of Supplementary Fig. 8). We also observed a left-broadening and a red-shift up to 3 $cm^{-1}$ as the laser power increases above threshold, which is particularly evident for the $A_{1g}$ peaks at 504 $cm^{-1}$ and 732 $cm^{-1}$ associated with oxygen breathing modes[54] (Fig. 3f-g). This is consistent with a transient laser heating-induced re-crystallization of YIG in a strained state, characterized by less energetic vibrational modes[55], and enhanced perpendicular magnetic anisotropy. Notably, strain-induced anisotropy is one of the most effective methods for establishing PMA in YIG thin films. In fact, both experimental and theoretical studies demonstrated how a tensile in-plane strain can generate a large magnetoelastic anisotropy in negative magnetostrictive materials like YIG[56–58]. While this is typically achieved through global processes, such as post-growth annealing, doping, or lattice mismatch with the substrate, our findings indicate that strain engineering via direct laser writing can also play a significant role.

Besides, the optical reflectance spectra in Fig. 3d show higher optical absorption below 450 nm, indicating the formation of oxygen vacancies upon irradiation[59,60], which may give rise to the strained state[59,61] and contribute to the perpendicular magnetic anisotropy by enhancing the magnetostriction coefficient[58]. Similar reflectance oscillations are instead observed above 450 nm, indicating only minor changes of the optical properties in the visible range (see also optical image in Fig. 3c). Noteworthy, the patterning UV laser wavelength (405 nm, violet arrow) is efficiently absorbed by YIG within the first hundreds of nanometres, allowing for an effective heat confinement in the vertical direction. This is the key mechanism enabling the three-dimensional tunability as a function of the laser power, as confirmed by our finite-elements methods modelling of the laser-YIG interaction discussed in Supplementary Note 2, Supplementary Fig. 1,2.

In order to investigate the stability of the laser-driven modifications and the role of strain and oxygen stoichiometry, we took MFM images of patterned squares before and after performing a rapid thermal annealing at 850°C in oxygen atmosphere. Supplementary Fig. 9 highlights that the new strained phase is robust against conventional thermal treatments, confirming the presence of a permanent structural modification upon laser irradiation above threshold. On the other hand, we observed that sub-threshold modifications can be reversed via thermal annealing, making the patterning fully reconfigurable under these conditions and establishing a promising platform for future investigations.

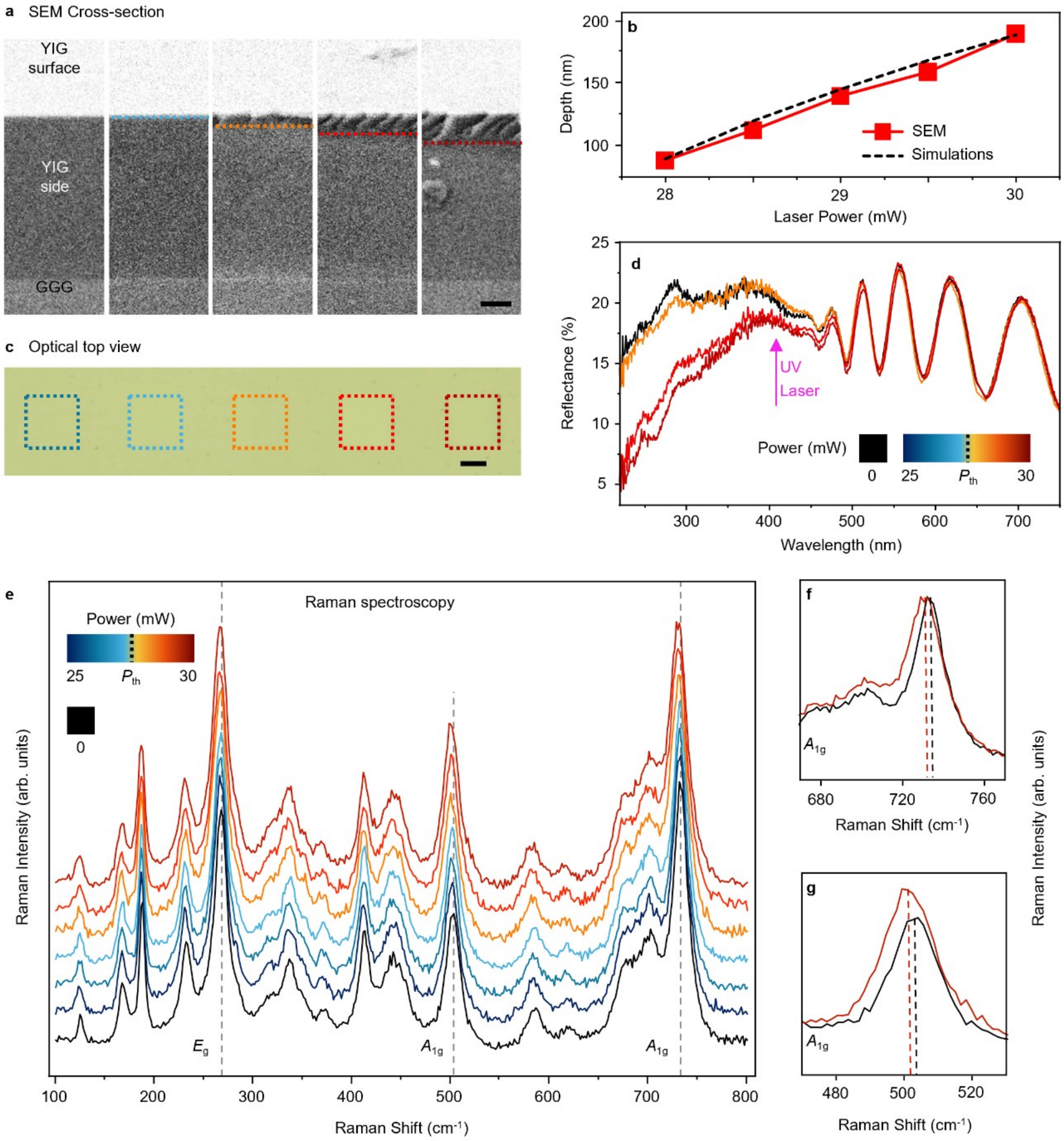

**Figure 3. Structural and optical characterisation. a,** Tilted cross-sectional Scanning Electron Microscopy images showing a tunable increase of the thickness of the modified volume with the laser power, enabling three-dimensional nanopatterning capabilities. The YIG surface appears smooth, while the sidewall roughness in the modified volume suggests strain relaxation at the edge. Scale bar: 200 nm. **b,** Experimental data measured from images in panel **a** and simulated values (see Supplementary Fig. 2) of the modified depth as a function of the laser power. The plot reveals a nearly linear relation which allows for continuous depth-control. **c,** Optical images of the irradiated areas, showing no sizeable optical contrast within the patterns at all powers. Scale bar: 25 μm **d,** Spectral reflectance as a function of the irradiation power. In the YIG transparency window above $\lambda = 450$ nm, no sizable difference is observed between the irradiated and pristine regions. The oscillatory behaviour due to thin film interference is consistent with a 1 μm-thick film. Below 450 nanometres, in the YIG absorption band, the regions irradiated above $P_{th}$ display lower reflectivity, indicating an enhanced optical absorption upon irradiation, suggesting the creation of oxygen vacancies. The 405 nm UV writing laser (pink) is efficiently absorbed by YIG. **e-g,** Raman spectra of the YIG film as a function of the irradiation power, and corresponding zoomed in view of the $A_{1g}$ (**f, g**) peak regions. The characteristic modes of single crystal YIG are present at all powers, with a minor left-side broadening and red-shift especially of the $A_{1g}$ peaks, above the threshold power, indicating a strained system.

**Tailoring spin waves**

To explore the potential of three-dimensional YIG nanopatterning in magnonics, we first studied the spin-wave properties of uniformly irradiated regions via Brillouin Light Scattering microscopy (micro-BLS) (Fig. 4a). An external magnetic field $H_{ext}$ was applied parallel to the microwave antenna used for exciting spin waves in the Damon-Eshbach (DE) configuration, where the spin-wave wavevector $k$ lies in the plane of the film, and is perpendicular to the magnetic field (Fig. 4b). Fig. 4c shows micro-BLS spectra as a function of the excitation frequency for different $H_{ext}$. In the pristine YIG (black curves), the BLS spectra feature a single intense peak and a weaker shoulder at lower frequencies. In the regions irradiated below threshold, i.e. up to 27 mW (blue curves), the BLS spectra remain nearly unchanged, in agreement with the static magnetic characterisation. On the contrary, above the threshold power (orange and red curves), the BLS spectra undergo a substantial evolution. In particular, new, less intense peaks appear at low frequencies, and the most intense peak first slightly broadens and shifts (at 28 mW), and then splits into a double peak (at 29 mW).

Hence, to better interpret the experimental results, we simulated the spin-wave dispersion for $\mu_0 H_{ext} = 125$ mT (see also Methods). In the pristine YIG (Fig. 4d), the DE mode (thick black curve) has an almost uniform spatial profile through the film thickness at $k = 0$ (Fig. 4g). Several perpendicular standing spin-wave modes are also present, due to the rather large thickness of the film (see thermal BLS in Supplementary Figure 10). The calculated dispersion curve is in good agreement with the micro-BLS experiments, where the ~ 6 GHz peak (black arrow) corresponds to the (DE) mode.

Similarly, for simulating the irradiated samples, we considered the same bicomponent system as in the static simulations of Fig. 2d-f, where the top modified volume of thickness $d$, retrieved from SEM measurements, has an enhanced perpendicular magnetic anisotropy (see Methods). The simulations (Fig. 4e, f) reveal significant changes of the spin-wave dispersion, which nicely explain the experimental features of Fig. 4c, where color-coded arrows indicate the peaks corresponding to the simulated modes.

First, new dispersive modes at low frequency appear, whose behaviour and number changes significantly with $d$. Their amplitude profiles (Fig. 4h, j-k) show that they are mainly localized in the top modified volume, rapidly vanishing through the thickness. In particular, for both $d$ = 100 nm and $d$ = 200 nm, the lowest frequency modes can be interpreted as the DE mode of the top volume. The other mode, only present for $d$ = 200 nm, can be considered as the first perpendicular standing spin-wave mode, featuring one node through the top volume thickness (Fig. 4k). Interestingly, the high frequency mode (the most intense peaks in Fig. 4c) significantly shifts its localization towards the bottom (Fig. 4i, l) and can be considered as the DE mode of the bottom non-modified volume.

Secondly, the marked difference between the positive ($k > 0$) and negative ($k < 0$) branches of the dispersions indicate a strongly non-reciprocal spin-wave propagation[62]. This is induced by the different out-of-plane anisotropy in the pristine bottom layer and the modified top layer, and the effect becomes more pronounced as the thickness of the modified volume increases (Fig. 4d-f). Accordingly, the presence of double peaks in Fig. 4c can be ascribed to the complex non-reciprocal dispersion and to the hybridisation with volume modes in the patterned regions.

These same phenomena also hinder an accurate estimation of the damping parameter in both pristine and patterned regions via micro-BLS measurements of the decay length of spin waves excited by RF antennas. In particular, the coexistence and interaction of Damon-Eshbach and perpendicular standing spin wave modes affect both the excitation efficiency of the antenna and the spatial mode profile. As a result, the observed decay of the spin-wave amplitude is strongly influenced by multiple overlapping modes with different propagation characteristics. This makes it difficult to reliably extract the intrinsic Gilbert damping from the measured decay length.

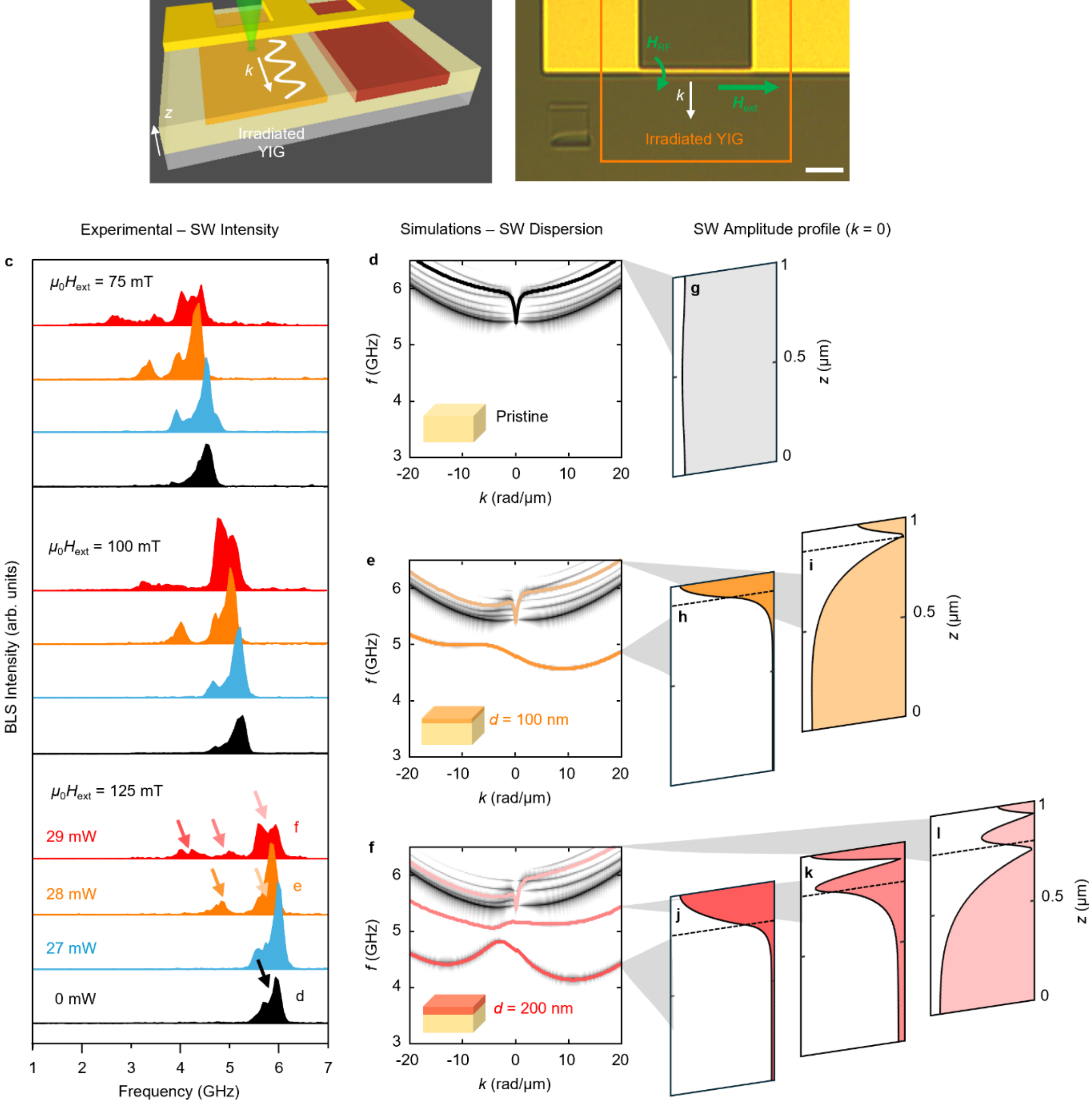

**Figure 4. Three-dimensional control of spin-wave dispersion and localization. a,** Sketch of the experimental setup. Extended square areas are irradiated with different uniform laser power. Radiofrequency antennas are fabricated on top of the irradiated regions for exciting spin waves via an oscillating magnetic field $H_{RF}$. The local spin-wave spectra are acquired via Micro-Brillouin Light Scattering using a focused laser, in presence of a static magnetic field $H_{ext}$ applied parallel to the antenna. **b,** Optical image of the sample. Scale bar: 10 μm. **c,** Micro-BLS spectra for different irradiating laser power, measured for different values of external field $H_{ext}$. Below threshold (in blue), the spectra are nearly unchanged with respect to the pristine sample (in black). Above threshold (in orange, red), additional low-frequency peaks appear which rapidly evolve increasing the laser power. **d-f,** Micromagnetic simulations of the spin-wave dispersions with $H_{ext}$ = 125 mT in the non-patterned film (**d**), for a 100 nm thick modified volume (**e**) and for a 200 nm thick modified volume (**f**). The corresponding experimental peaks are indicated by color-coded arrows in (**c**). **g-l,** Simulated spin-wave amplitude profiles through the thickness of the film, for the modes highlighted in panels **d**-**f** at $k$ = 0. The simulations show the appearance of additional non-reciprocal modes upon irradiation, localized within the modified volume, and finely controllable with the irradiation power.

However, it is important to note that neither a larger Gilbert damping nor a slight topographic dip in irradiated areas would account for such features, that can ultimately be attributed to the enhanced PMA (see Supplementary Fig. 11). This demonstrates that laser-written three-dimensional anisotropy profiles enable the effective control over the spin-wave 3D spatial localization, magnonic bandstructure and non-reciprocity with potential application in vertically integrated magnonic devices.

## Direct-write magnonic crystals

We harnessed these capabilities to directly fabricate proof-of-principle magnonic crystals based on non-uniform perpendicular magnetic anisotropy.

As illustrated in Fig. 5a, a first experimental demonstration is provided by spin-wave spectroscopy performed with a Vector Network Analyzer (VNA) on a magnonic crystal positioned between two RF stripline antennas, separated by 100 μm, and used for the excitation and detection of spin waves under an in-plane magnetic field $H_{ext}$ = 95±5 mT in the DE configuration (see Methods). The magnonic crystal is made of three laser-written lines, each 1.5 μm-wide and separated by $l$ = 25 μm (MFM image shown in Fig. 5b). This periodic modulation is designed to induce Bragg reflection, thereby modifying the spin-wave transmission spectrum. The results are presented in Fig. 5c. While both the pristine material and the magnonic crystal display a similar overall transmission window (~ 4–5 GHz) with comparable maximum intensity, the magnonic crystal clearly exhibits additional transmission minima at discrete frequencies, corresponding to wavevectors $k = n\pi/l$ derived from Bragg's law[39]. The bandgap formation confirms the successful realization of a magnonic crystal through direct laser writing. In contrast, the dips observed in the pristine spectrum are attributed to the hybridization of the DE spin-wave mode with perpendicular standing spin-wave (PSSW) modes[63].

Furthermore, we fabricated a second magnonic crystal consisting of a square lattice of circular dots, single-shot irradiated with a 27.5 mW laser power (see Methods). The dots diameter is $a$ = 600 nm, while the lattice constant is $l$ = 2 μm. Micro-BLS in the DE configuration (see Fig. 5d), with a 75 mT external magnetic field, was used to acquire 2D maps of the spin-wave intensity of the two main detected modes at 4.0 GHz (Fig. 5e) and 2.6 GHz (Fig. 5f). The spatial localization of both modes is strongly affected by the irradiated dots, similarly to bicomponent magnonic crystals[64–67]. In particular, the 4.0 GHz mode (Fig. 5e) extends within the horizontal channels comprised between the irradiated dots, and has a sizeable intensity within the dots as well. Conversely, the 2.6 GHz mode (Fig. 5f) extends along vertical channels and its intensity spatially oscillates with the same periodicity of the antidot lattice, with an opposite phase in adjacent channels. The maps of both modes are in good agreement with the simulations shown in Fig. 5g, i (see Methods).

Remarkably, the confinement of the magnetic anisotropy patterns induces non-uniform spin-wave profiles across the thickness. In particular, the simulated cross-sections (Fig. 5h, j) show that, in the channels containing (not containing) the dots, both modes extend mainly in the bottom (top) volume. This gives rise to a spatial oscillation of the spin-wave intensity throughout the volume that, combined with the in-plane spatial modulation typical of 2D magnonic crystal, generates fully three-dimensional spin-wave mode profiles. In perspective, the possibility of controlling point-by-point the modified volume thickness allows to realize, in a single step, YIG-based complex "multicomponent" crystals or grayscale magnonic metamaterials where the spin-wave properties are tailored via smooth three-dimensional magnetic anisotropy landscapes.

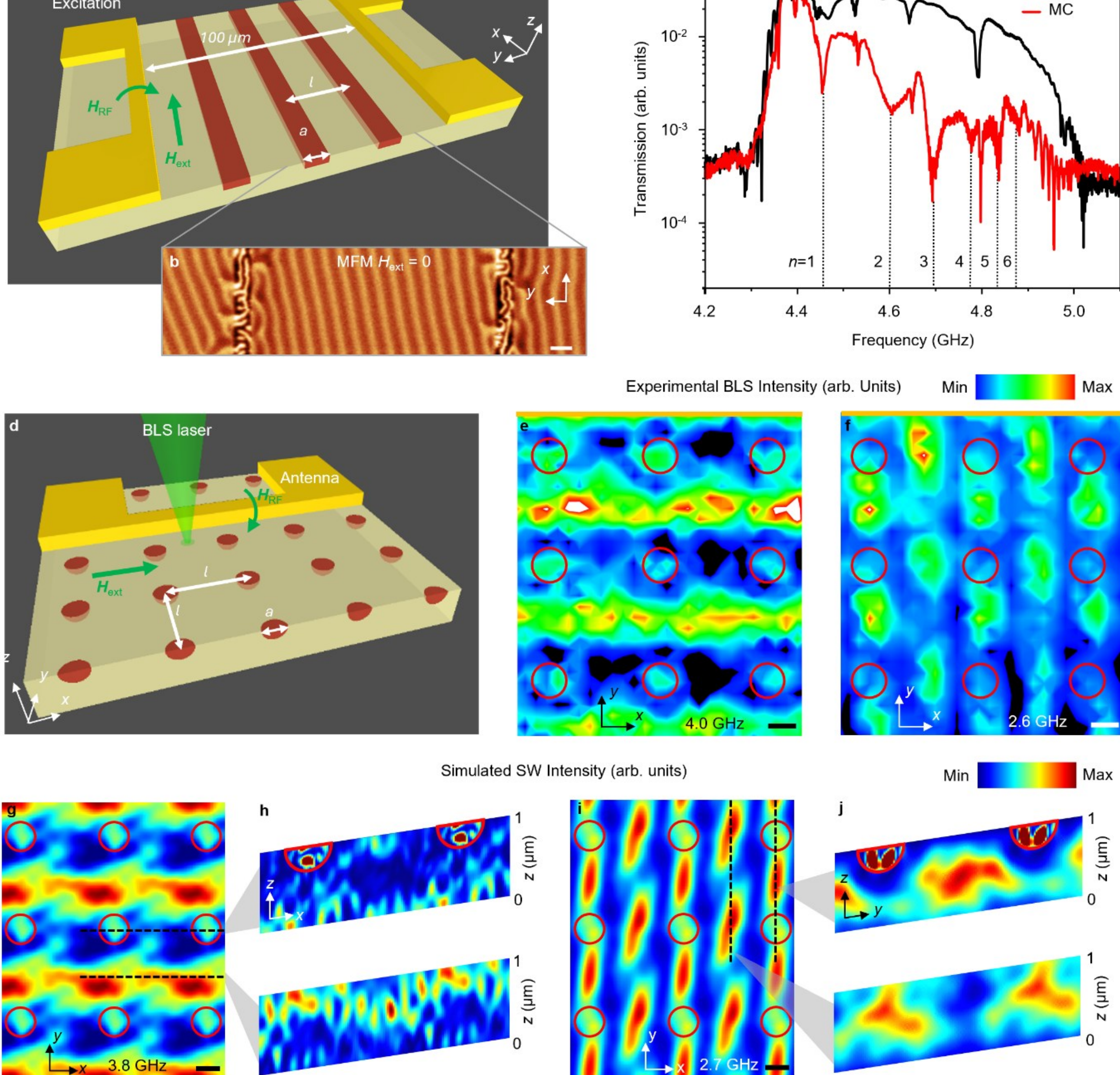

**Figure 5. Direct-write magnonic crystals**. **a**, Schematic of the propagating spin-wave spectroscopy experiment. A magnonic crystal (MC) consisting of three patterned lines having a width of $a$ = 1.5 µm and a periodicity of $l$ = 25 µm acts as a Bragg reflector for spin waves. Excitation and detection are performed with two stripline antennas separated by 100 µm, applying a 95 mT in-plane static magnetic field in the DE configuration. **b**, MFM image of the magnonic crystal showing two lines with narrower stripe domains due to the laser-induced PMA enhancement. Scale bar: 1.5 µm. **c**, Spin-wave transmission spectra of the pristine material (black curve) and the magnonic crystal (red curve). While the transmission window remains unchanged, the formation of band gaps (vertical dotted lines), consistent with Bragg reflection is observed in the magnonic crystal. **d**, Schematic of the Brillouin light scattering experiment. A magnonic crystal with a periodic circular anti-dot lattice is realized via single-shot irradiation of the red areas. The dot diameter is $a$ = 600 nm and the lattice constant is $l$ = 2 µm. A stripline antenna is fabricated on top of the magnonic crystal for spin-wave excitation. The spin-wave intensity is mapped via micro-BLS in a 75 mT static magnetic field applied along $x$. **e, f,** Micro-BLS spin-wave intensity maps showing horizontal spin-wave channels between the dots in the 4.0 GHz mode (**e**), and vertical spin-wave channels with a spatially oscillating intensity in the 2.6 GHz mode (**f**). **g, i,** Simulated spin-wave intensity maps corresponding to measured spin wave modes in (**e, f**). **h, j**, Cross-section view of the simulated spin-wave intensity through the thickness of the film, extracted along the dashed lines in (**g, i**). In both cases, the three-dimensional magnetic properties profiles determine a non-uniform spin-wave intensity and localization within the volume of the film. Scale bars: 500 nm.

## DISCUSSION

We demonstrated the creation of three-dimensional nanoscopic modulations in the magnetic anisotropy of crystalline Yttrium Iron Garnet, via direct irradiation with a continuous-wave UV laser in ambient conditions, preserving the crystalline quality. The 3D tunability and spatial resolution arises from linear absorption mechanisms, which enable to confine and finely modulate the volume subject to the heat-triggered phase transition with an experimental lateral minimum feature size of 100 nm and continuous depth-control. Furthermore, we demonstrated that such modulations can be induced at the buried GGG/YIG interface, as well, by irradiating through the transparent GGG substrate. In parallel, direct modification via other mechanism such as non-linear multiphoton absorption with ultrafast lasers could be investigated.

We exploited this capability for tailoring the static and dynamic magnetic properties, realizing 3D magnonic crystals and nanostructures. In such systems, the spin texture, spin-wave band structure and spatial localization are governed by the vertical extension of the modified volume. This opens the way to a new class of nanomaterials requiring a three-dimensional tuning of the magnetic properties, enabling novel nanoscale, low-loss, and reconfigurable magnonic devices. The technique is also highly complementary to other methods that can induce temporary changes in magnetic properties to excite and control spin waves. In principle, these approaches could be combined to selectively create permanent three-dimensional patterns alongside dynamic regions within the same YIG system.

The technique is straightforward, rather inexpensive and potentially easily transferable towards industrial applications. We believe it represents a new valuable tool enabling the single-step realization of nanodevices exploiting spatially varying magneto- and magneto-optical properties, overcoming some of the hurdles of conventional nanofabrication, and opening the path towards three-dimensional metamaterials and devices based on crystalline magnetic garnets. Beyond magnetism, a similar concept could be applied to other crystalline complex oxides, where new functionalities would arise from tailored three-dimensional modulations of their structural properties.

## METHODS

### Samples and antennas fabrication

The 1 μm-thick crystalline YIG film used in this study was grown by liquid phase epitaxy (LPE) on 500 μm-thick GGG (111) substrate (by Matesy GmbH). For the micro-BLS experiments, stripline antennas having a width of 1 μm were fabricated using maskless photolithography and electron beam lithography, followed by thermal evaporation of Cr(10 nm) / Au (120 nm) and lift-off.

### Direct Laser Writing

The laser patterning was performed with the NanoFrazor Explore by Heidelberg Instruments Nano AG. The system is equipped with a continuous-wave (CW) semiconductor diode laser emitting at 405 nm with a nominal spot size of 1.2 μm, which is focused on sample surface through a 20x objective. Each pattern is divided into 50 nm × 50 nm pixels; accordingly, the piezo stage moves the sample in a raster scan fashion with constant velocity given by the ratio between pixel size and pixel time (50 μs). Target pixels are sequentially exposed for a duration of 20 μs. The pattern occurs only where the temperature reached by the film upon laser irradiation allows for the selective melting and recrystallization of YIG. By exploiting this threshold mechanism and the Gaussian intensity profile at low laser power, super-resolution can be achieved, well below the beam's spot size. To calculate the threshold laser power density, we considered an ideal gaussian beam with beam radius at the surface $w_s = 1.3$ μm. This value is higher than the beam waist $w_0 = 0.6$ μm, as during patterning the laser was focused ≈ 5 μm below the surface, resulting in a larger radius at the surface, according to the simulations in Supplementary Note 2. The threshold power density at the surface is therefore estimated as $I_{th} = P_{th} / (\pi w_s^2) = 0.52 \cdot 10^6$ W/cm$^2$.

### Atomic and Magnetic Force Microscopy

MFM images were taken using Park Systems NX10 AFM in lift mode with PPP-LM-MFMR magnetic probes by Nanosensors. The corresponding topographic maps were jointly obtained during the same measurement. Data processing included polynomial background subtraction and correction of scan line artifacts and mark scars. MFM measurements as a function of the external magnetic field were performed using a Caylar Magnetic Field Module system integrated with the Park NX10.

### Magneto-optic measurements

Hysteresis loops were measured by a custom-made setup exploiting the Magneto-Optic Kerr Effect (MOKE). The sample was placed between the poles of an electromagnet, which generates a magnetic field oriented either parallel (longitudinal configuration) or perpendicular (polar configuration) to the sample surface. A 405 nm laser light is p-polarized before impinging on the sample, with an angle of 50° with respect to the normal direction and with a spot size of about 30 μm. The reflected beam passes through a polarizer with an axis tilted by 7° with respect to the s polarization (analyser) and it is collected by a photodiode. Additionally, the laser light was modulated by an optical chopper and then demodulated using a lock-in amplifier (Zurich Instruments HF2LI) in order to get rid of low frequency noise. Final hysteresis loops are the average of five subsequent measurements in which each data point was acquired ten times. As for data treatment, the contribution of the hysteretic behaviour of the electromagnet was accounted and a linear trend beyond saturation was subtracted, and finally normalized. The saturation field was determined as the magnetic field at which the normalized magnetization attains 95% of its saturation value, with error bars given by the magnetic field step size.

### Scanning Electron Microscopy

After exposing the sample with different laser power, it was flipped and the substrate was cut with a dicing saw (Disco DAD341) in correspondence of the irradiated areas, keeping a working distance of 50 μm from plane level (much higher that YIG thickness equal to 1 μm). In this way, the sample was cleaved along the line with minimum damage. Scanning Electron Microscopy (Leo/Zeiss 1525) images were then taken using the Everhart–Thornley secondary electron detector with an accelerating voltage of 3 kV and beam current of 130 pA. A tilt angle of 20° allowed for both surface and cross-section visualization of the sample.

**Raman Spectroscopy**

Raman spectra were recorded with a Renishaw InVia micro-Raman spectrometer. The employed light source is a 457 nm Argon ion laser, focused through a 50x objective, providing an actual laser power of 1.3 mW on sample surface. A single spectrum contains 444 evenly distributed data points corresponding to Raman shifts from 100 $cm^{-1}$ to 800 $cm^{-1}$. The acquisition time was set to 10 s and each final spectrum is the sum of 20 repetitions to increase the signal to noise ratio. Data curation comprehends linear background subtraction and normalization between zero and the highest peak at 732 $cm^{-1}$.

**BLS measurements**

Micro-BLS measurements was carried out by focusing a single-mode solid-state laser (with a wavelength of 532 nm) at normal incidence onto the sample surface using an objective with numerical aperture of 0.75 and a working distance of 4.7 mm, giving a spatial resolution of about 250 nm. A (3+3)-pass tandem Fabry-Perot interferometer was used to analyse the inelastically scattered light. A nanopositioning stage allowed us to position the sample with a precision down to 10 nm on all three axes. A DC/AC electrical probe station ranging from DC up to 20 GHz with frequency steps of 50 MHz was used for spin-wave pumping. A stripline antenna with a width of $w$ = 1 μm was used to efficiently excite spin waves with wavevectors up to $k = \pi/w > 2.5$ rad/μm. Fig. 4c shows micro-BLS spectra taken at a distance of about 3 μm from the antenna. Fig. 5c, e show 6 x 8 $\mu m^2$ micro-BLS maps acquired with a step of 200 nm in $x$ and $y$.

**Propagating spin-wave spectroscopy**

Spin-wave spectroscopy was performed with the Rhode & Schwarz ZNA43 Vector Network Analyzer (VNA) and a probe station equipped with a quadrupole electromagnet. The intensity of the transmitted signal was extracted as the modulus of the $S_{21}$ parameter, which represents the measured transmission of spin waves from the antenna 1 to the antenna 2, from which a reference signal measured for $H_{ext} = 0$ was first subtracted: $|\Delta S_{21}(f)| = |S_{21}(f) - S_{21,\mathrm{ref}}(f)|$. The power of the RF excitation signal was set to -10 dBm and we acquired 20000 points between 1 GHz and 9 GHz.

**Micromagnetic Simulations**

Micromagnetic simulations of remanence states, magnetic hysteresis and magnetization dynamics were carried out by solving the Landau–Lifshitz equation, using the GPU-accelerated software Mumax[3]. Different types of simulations were performed in order to understand the behaviour of the material before and after the laser irradiation. Common magnetic parameters are: saturation magnetization $M_s$ = 140 kA/m; exchange stiffness A = 4 pJ/m; anisotropy constant of the pristine film $K_U$ = 0.2 $kJ/m^3$; anisotropy constant of the modified layer $K_U$ = 6 $kJ/m^3$. A thorough description of simulation parameters and procedure is given in Supplementary Note 1.

**Additional structural, optical and magnetic characterisation**

Reflectance spectra were acquired with a Filmetrics F54-XY-UVX system through a 10x objective.

Magneto-Optic Kerr Effect images were acquired with a custom-made setup based on a microscope equipped with a high power stabilized LED white light source, a 20x objective and a CCD camera. The measurements exploit the polar MOKE effect, which is sensible to the out-of-plane component of the magnetization, and were performed while sweeping an external in-plane magnetic field generated by an electromagnet.

The X-Ray Diffraction θ-2θ scan was performed with a Rigaku SmartLab XE X-Ray Diffractometer with Cu-Kα radiation ($\lambda$ = 1.5406 Å).

Electron Backscattered Diffraction inverse Pole Figure maps were obtained using a Zeiss Sigma 500 Field Emission Scanning Electron Microscope equipped with the Oxford C-NANO Electron Backscattered Diffraction detector.

## DATA AVAILABILITY

The data generated in this study have been deposited in the Zenodo database under accession code 10.5281/zenodo.17132490 [https://doi.org/10.5281/zenodo.17132490].

## ACKNOWLEDGMENTS

This work was partially performed at PoliFab, the microtechnology and nanotechnology center of the Politecnico di Milano. The authors thank PoliFab staff, particularly Elisa Sogne, Marco Asa, Andrea Scaccabarozzi and Stefano Bigoni for help with the SEM, XRD and optical characterisation, and Simone Finizio for help with fabrication. E.A. acknowledges funding from the European Union's Horizon 2020 research and innovation programme under grant agreement number 948225 (project B3YOND) and from the FARE programme of the Italian Ministry for University and Research (MUR) under grant agreement R20FC3PX8R (project NAMASTE). E.A., S.T. and G.C. acknowledge funding from the European Union – Next Generation EU – "PNRR – M4C2, investimento 1.1 – "Fondo PRIN 2022" – TEEPHANY– ThreEE-dimensional Processing tecHnique of mAgNetic crYstals for magnonics and nanomagnetism ID 2022P4485M CUP D53D23001400001". D.P. acknowledges funding from the European Union – Next Generation EU – "PNRR – M4C2, investimento 1.1 – "Fondo PRIN 2022" – PATH – Patterning of Antiferromagnets for THz operation id 2022ZRLA8F – CUP D53D23002490006" and from Fondazione Cariplo and Fondazione CDP, grant n° 2022-1882. P. P., A. D. G., M. C., F. M., R. B., and S. T. acknowledge funds from the EU project MandMEMS, grant 101070536. R. B. acknowledges funding from NextGenerationEU, PNRR MUR – M4C2 – Investimento 3.1, project IR_0000015 – "Nano Foundries and Fine Analysis – Digital Infrastructure (NFFA–DI)", CUP B53C22004310006. L. C. M., M. M., R. S. and S.T. acknowledge financial support from NextGenerationEU National Innovation Ecosystem grant ECS00000041–VITALITY (CUP B43C22000470005 and CUP J97G22000170005), under the Italian Ministry of University and Research (MUR). D.B. and P. P. acknowledge funding by the Deutsche Forschungsgemeinschaft (DFG, German Research Foundation) - TRR 173/3 - 268565370 Spin+X (Project B01) and by the ERC Grant No. 101042439 'CoSpiN'.

## AUTHOR CONTRIBUTIONS

These authors contributed equally: Valerio Levati and Matteo Vitali. V.L., M.V. and D.G. performed the laser nanopatterning. V.L., M.V. and N.P, performed the static magnetic characterisation. M.M. and S.T. performed the BLS measurements and data analysis. M.V., L.C.M., R.S. performed the micromagnetic simulations. V.L., M.V., A.D.G., I.B., M.P., P.F., L.R., N.L., F.M., R.B., performed the structural and optical characterisation. V.L., M.V., V.R. and A.L.B. performed the Raman measurements. I.B., M.P., P.F. and D.P. fabricated the spin-wave antennas. V.L., M.V. and M.C., performed the spin-wave spectroscopy measurements. M.V. developed the FEM laser heating model. V.L., M.V., S.T. and E.A. wrote the manuscript aided by D.B., P.P., G.C., R.O. and D.P. with contributions from all the authors. All authors contributed to discussion and data interpretation. S.T., D.P. and E.A. designed and supervised the research.

## COMPETING INTERESTS

The authors declare no competing interests.

## SUPPLEMENTARY INFORMATION FOR:

# Three-dimensional nanoscale control of magnetism in crystalline Yttrium Iron Garnet

Valerio Levati, Matteo Vitali, Andrea Del Giacco, Nicola Pellizzi, Raffaele Silvani, Luca Ciaccarini Mavilla, Marco Madami, Irene Biancardi, Davide Girardi, Matteo Panzeri, Piero Florio, Maria Cocconcelli, David Breitbach, Philipp Pirro, Ludovica Rovatti, Nora Lecis, Federico Maspero, Riccardo Bertacco, Giacomo Corrielli, Roberto Osellame, Valeria Russo, Andrea Li Bassi, Silvia Tacchi, Daniela Petti, Edoardo Albisetti

**Supplementary Note 1:** Detailed description of micromagnetic simulations

### Remanence states

To simulate the irradiated patterns, a bicomponent system was considered, with the top layer representing the irradiated material with an increased anisotropy constant of $K_U$ = 6 kJ/m$^3$. This approximation is justified by the fact that the patterning is performed through a threshold mechanism of melting and recrystallization; the thermal gradient induced by the laser does not gradually modulate the magnetic properties, but rather determines the depth of the patterned region.
The geometry of the simulation is a long stripe (10.24 μm x 25 nm x 1 μm) discretized in cubic cells with a 5 nm edge. Periodic boundary conditions were considered in the in-plane directions. For the pristine material, the following parameters were assumed: saturation magnetization of $M_s$ = 140 kA/m, uniaxial OOP anisotropy constant of $K_U$ = 0.2 kJ/m$^3$ and an exchange stiffness of $A$= 4 pJ/m. To simulate the irradiated patterns, a bicomponent system was considered, with the top layer representing the irradiated material with an increased anisotropy constant of $K_U$ = 6 kJ/m$^3$. The system was initially saturated along the y-axis with a high magnetic field. The magnetization was allowed to precess, considering a damping of $\alpha = 0.03$, then the system was relaxed and the field was decreased. The procedure was repeated until the field value reached zero. Figures 2 d-f represent relaxed states for an external field value of 10 Oe (1mT). During each iteration, for high field values, the magnetization precession time was 1 ns, and the field was reduced by steps of $\Delta H$ = 40 Oe and $\Delta H$ = 100 Oe when simulating the pristine and bicomponent stripe respectively. For the pristine simulations, the precession time was increased to 20 ns and the field step was reduced to $\Delta H$ = 2 Oe for fields below $H$ = 60 Oe. For the bilayer stripe, the precession time was set to 10 ns for field values close to the saturation field and 20 ns for fields below $H$ = 60 Oe, keeping a field step of $\Delta H$ = 10 Oe.

### Dispersion curves and spatial profiles for continuous films

The dispersion relation curves of the continuous films were numerically calculated with a mesh size of 2048×2×100, where each cell measures 15×100×10 nm³. Periodic boundary conditions are applied in the x- and y-directions. The magnetic parameters for pristine YIG were: saturation magnetization $M_s$ = 140 kA/m, exchange stiffness $A$ = 4 pJ/m, and out-of-plane uniaxial anisotropy constant $K_U$ = 0.2 kJ/m$^3$. For the irradiated part of the film, the only parameter change is the uniaxial anisotropy constant, which is set to $K_U$ = 6 kJ/m$^3$ as in the “remanence state” simulations.
A static magnetic field of $H$ = 1250 Oe was applied along the negative y-direction, while spin waves were excited applying a dynamic magnetic field along the out-of-plane direction over the entire thickness, but within a small central portion of the film, which is 90 nm wide. The dynamic field in the time domain is a sinc pulse with maximum frequency $f_0$= 30 GHz and amplitude $b$ =10 mT. The simulations are performed for 50 ns, and the $m_z$component of the magnetization vector is extracted every 50 ps. The dispersion relations are calculated by averaging the absolute value of the Fast Fourier Transform (FFT) in space and time of the $m_z$ component along the thickness. The spatial profiles, shown in Figure 4g-l, are obtained from dedicated simulations with a reduced mesh size of 8x2x100, where the dynamic field is applied over the entire simulated area to excite the spin waves with zero wave vector. In Figure 4, the real part of the temporal FFT of the $m_z$ component is shown, extracted from the first three peaks in the spectrum obtained by averaging the absolute value of the temporal FFT of the $m_z$ component.

### Spin waves intensity maps

For these simulations, a 2 μm x 2 μm x 1 μm structure was considered, discretized with 80 x 80 x 100 cells. Periodic boundary conditions in both in-plane directions were assumed. The following parameters were considered: $M_s$= 140 kA/m, $A$ = 4 pJ/m, uniaxial OOP anisotropy constant of $K_U$ = 0.2 kJ/m$^3$ and damping $\alpha$ = $10^{-4}$. In the centre of the top surface a dot structure was simulated as a bisected ellipsoid with an in-plane diameter of 600 nm and a thickness of 300 nm, with an increased anisotropy of $K_U$ = 6 kJ/m$^3$. This geometry was chosen to accurately model the physical shape of the modified region, which is wider at the sample surface and tapers with depth due to optical absorption. The system was initially saturated by a 75 mT in plane field with a one degree tilt respect to the x axis, then the whole structure was perturbed by a sinc-like impulse along the z axis, having the following expression

$$B_z = B_0(z)\frac{\sin\left(2\pi f(t-t_0)\right)}{2\pi f(t-t_0)},$$

$$B_0(z) = -\frac{B_0}{2}\ln\left(\frac{(y_0-L/2)^2+(z-d)^2}{(y_0+L/2)^2+(z-d)^2}\right),$$

with $t_0$ = 1 ns, $f$ = 15 GHz, $B_0$ = 5 mT, $y_0$ = 550 nm, $L$ = 1 μm, $d$ =1 μm. $B_0(z)$ represents the $z$ component of a Karlqvist field generated by a microstrip antenna of width $L$, parallel to the $x$ axis and placed on the top surface, evaluated at a constant distance of $y_0$ from the strip centre ($z=0$ is the bottom of the structure and $d$ is the thickness)[1]. The total simulation time was 40 ns and the magnetization maps were acquired every 50 ps. To reproduce the micro-BLS spin wave intensity maps of Figure 5g, i, the absolute value of the FFT transform of the $M_z$ component was averaged along the thickness of the system and its square was then plotted. Also, by employing the periodic boundary conditions, repetitions of the obtained frequency map were added at the sides of the output. The extended maps were convoluted with a gaussian with 250 nm FWHM to match the experimental results. To obtain the sections in Figure 5h, j, the square of the absolute value of the FFT elaborated $M_z$ was extracted from slices through the thickness of the structure and then extended along the in-plane direction employing the periodic boundary conditions.

### Out of Plane hysteresis simulations

To simulate the OOP hysteresis loops a 2.56 x 2.56 x 1 um$^3$ structure was considered, discretized with a 10 nm edge cubic cell, with periodic boundary conditions assumed in both in-plane directions. The system was initially saturated along the positive z-direction with a $B_z$ = 250mT field. Then, the field was swept twice between 250 mT and -250 mT, with steps of 10 mT. For each step, the magnetization was allowed to precess for few nanoseconds (2 ns for fields below 200 mT; 0.5 ns for other fields) considering a damping of $\alpha$ = 0.03, and then relaxed. When simulating the pristine material, the following parameters were considered for the whole structure: saturation magnetization $M_s$ = 140 kA/m , exchange stiffness $A$ = 4 pJ/m, OOP uniaxial anisotropy constant $K_U$ = 0.2 kJ/m$^3$. To simulate the irradiated patterns, a bilayer structure in which the top 100 nm or 200 nm of the structure were considered to have an increased anisotropy constant of $K_U$= 6 kJ/m$^3$ was considered. In Supplementary Figure 7b are represented the simulated hysteresis cycles, in which the magnetization is averaged over all cells for the pristine structure and over the top 100 nm or 200 nm cells for the bilayer simulations.

**Supplementary Note 2:** Modelling local laser heating in YIG

A COMSOL Multiphysics model was employed to simulate the heating of the YIG film which follows from the exposure to ultraviolet light during the patterning process. First, we calculated the properties of a gaussian laser beam propagating from air, through the YIG film, and in the GGG substrate. According to the Lambert-Beer law, we then considered the absorption of the optical power as the heating mechanism in the system. Finally, the heat flow was modelled to obtain information on the temperature profile in the YIG layer.

### Laser beam profile during propagation

We consider the intensity of a Gaussian beam as described by the following spatial profile

$$I(x,y,z) = \frac{2P_0}{\pi w^2(z)} e^{\left[-2\frac{x^2+y^2}{w^2(z)}\right]},$$

where $P_0$ is the laser power and $w(z)$ is the beam width as a function of the propagation coordinate $z$. The beam width is conventionally defined as the radius for which the intensity of the beam is lowered by a factor $1/e^2$ with respect to the maximum intensity. It is proportional to the beam waist $w_0$, which is the minimum radius of the laser beam at the focal point, i.e. the minimum radius. These quantities are linked by the following relations

$$w(z) = w_0\sqrt{1+\left(\frac{z-z_0}{z_R}\right)^2}, \qquad z_R = \frac{n\pi w_0^2}{\lambda},$$

where $z_0$ is the focus position, $z_R$ is the Rayleigh length, $n$ the index of refraction of the medium in which the beam propagates and $\lambda$ the free space wavelength of the laser.
The different indexes of refraction corresponding to the propagation across different materials entail a variation of the focus position and a different expression for the beam radius in each media.
Considering an infinite $xy$ plane, along the propagation direction $z$ we have air for $z > 0$, YIG for $-d < z < 0$, and GGG for $z < -d$. Here, the thickness of the YIG layer is $d = 1$ µm. Thence, the profile of the beam waist propagating in the system can be found using the subsequent expressions[2]

$$w_{VAC}(z) = w_0\sqrt{1+\left(\frac{z-z_0}{z_{R,VAC}}\right)^2}, \qquad z_{R,VAC} = \frac{\pi w_0^2}{\lambda};$$

$$w_{YIG}(z) = w_0\sqrt{1+\left(\frac{z-z_{0,YIG}}{z_{R,YIG}}\right)^2}, \qquad z_{R,YIG} = \frac{n_{YIG}\pi w_0^2}{\lambda}, \qquad z_{0,YIG} = z_0 n_{YIG};$$

$$w_{GGG}(z) = w_0\sqrt{1+\left(\frac{z-z_{0,GGG}}{z_{R,GGG}}\right)^2}, \quad z_{R,GGG} = \frac{n_{GGG}\pi w_0^2}{\lambda}, \quad z_{0,GGG} = z_0 n_{GGG} + d\frac{n_{GGG}}{n_{YIG}} - d.$$

In our case, $z_0 = -5.35$ µm, $\lambda = 405$ nm and $w_0$=0.6 µm, where $z_0$ can be interpreted as the focus position that the beam would have if it propagated only in free space, measured from the YIG-air interface. The refractive index $n_{YIG} = 2.806$ has been calculated from the real ($\varepsilon_1$) and imaginary ($\varepsilon_2$) part of the dielectric constants, as extracted from[3] with the following formula

$$n_{YIG} = (1/\sqrt{2})\sqrt{\varepsilon_1 + \sqrt{\varepsilon_1^2 + \varepsilon_2^2}},$$

while the refractive index $n_{GGG} = 2.015$ was retrieved from[4].
The profile of a laser beam propagating in the structure is plotted in Supplementary Figure 1.

### Laser beam attenuation

Given the absorption coefficient $\alpha$ of a medium, the optical attenuation through the system for a beam propagating along the negative $z$ direction is ruled by

$$\frac{dI}{dz} = \alpha I, \tag{1}$$

from which it follows the Lambert-Beer law $I = I_0 e^{-\alpha|z|}$, with $I_0$ constant. In particular, in our model the attenuated intensities can be written by multiplying the gaussian beam intensities by the exponential attenuation[5]:

$$I_{YIG} = \frac{2P_0(1-R_1)}{\pi w_{YIG}^2(z)} e^{\left[-2\frac{x^2+y^2}{w_{YIG}^2(z)}\right]} e^{-\alpha_{YIG}|z|}, \tag{2}$$

$$I_{GGG} = \frac{2P_0(1-R_1)(1-R_2)e^{-\alpha_{YIG}d}}{\pi w_{GGG}^2(z)} e^{\left[-2\frac{x^2+y^2}{w_{GGG}^2(z)}\right]} e^{-\alpha_{GGG}|z+d|}. \tag{3}$$

The YIG absorption coefficient was calculated as $\alpha_{YIG} = 4\pi\kappa/\lambda = 58.023\times10^3$ cm$^{-1}$, where $\kappa$ has been calculated from $\varepsilon_1$ and $\varepsilon_2$ extracted from[3] with the following formula

$$\kappa = \left(1/\sqrt{2}\right)\sqrt{-\varepsilon_1 + \sqrt{\varepsilon_1^2 + \varepsilon_2^2}}.$$

The absorption coefficient of GGG, $\alpha_{GGG} = 0.217$ cm$^{-1}$, was instead retrieved from[4]. Note that we assume the absorption coefficients to be independent of laser power and temperature.

The Fresnel coefficients $R_1$ and $R_2$ for normal incidence take into account the transmission at the interfaces, and are calculated according to

$$R_1 = (n_{YIG} - 1)^2/(n_{YIG} + 1)^2,$$
$$R_2 = (n_{GGG} - n_{YIG})^2/(n_{GGG} + n_{YIG})^2.$$

The back-reflection form the YIG/GGG interface was neglected without affecting the resulting optical power profile.

### Heat diffusion and temperature gradients

The system was modelled as a 250 x 250 μm$^2$ YIG/GGG bilayer. The YIG film is located between $z = 0$ and $z = -d$, with $d = 1$ μm is the film thickness, while the GGG substrate ($-d-d_{sub} < z < -d$) is modelled with a lower thickness ($d_{sub} = 100$ μm) compared to the real sample for reducing the computational time. This can be done since the beam power rapidly extinguishes through the thickness. The heat conduction in the system was modelled using the heat flow equation

$$\rho C_p \frac{\partial T}{\partial t} - \nabla \cdot (k\nabla T) = Q,$$

where $\rho$ and $C_p$ are the density and the specific heat of the material, $T$ is the absolute temperature, k is the thermal conductivity and $Q$ is the heat source. The parameters of YIG are already present in the COMSOL library, while those of GGG were retrieved in the literature: $k_{GGG} = 7.4$ W m$^{-1}$ K$^{-1}$, $\rho_{GGG} = 7080$ kg m$^{-3}$ and a constant $C_{P,GGG}(293.15\text{ K}) = 367$ J kg$^{-1}$ K$^{-1}$[6].

In our case, the heat was considered to originate solely from the absorption of light, according to equations (2) and (3). Hence, the energy sources for the two layers were defined as in[7]:

$$Q_{YIG} = \frac{2P_0(1-R_1)}{\pi w_{YIG}^2(z)} e^{\left[-2\frac{x^2+y^2}{w_{YIG}^2(z)}\right]} \alpha_{YIG} e^{-\alpha_{YIG}|z|} \quad (4)$$

$$Q_{GGG} = \frac{2P_0(1-R_1)(1-R_2)e^{-\alpha_{YIG}d}}{\pi w_{GGG}^2(z)} e^{\left[-2\frac{x^2+y^2}{w_{GGG}^2(z)}\right]} \alpha_{GGG} e^{-\alpha_{GGG}|z+d|}$$

The temperature was set to the ambient value of 293.15 K. The same value was fixed for the bottom of the GGG layer while natural external convection was considered above the sample surface. Supplementary Figure 2 shows the result of simulations run for 20 μs for laser power ranging from 27 mW to 30 mW. Noteworthy, the optical power is almost completely absorbed within the YIG layer, with the intensity rapidly dropping in the first 200 nm. As a consequence, only the YIG film undergoes a significant heating process. In particular, for laser powers lower than 27.5 mW (below the threshold power) the maximum temperature does not exceed the melting point, i.e. 1828 K; while for larger powers the melting temperatures is reached inside the YIG layer with a tunable depth increasing with optical power. This gives an insight on the physical effect of laser heating, which thus involves the transient melting of the topmost layer of the material, and its subsequent recrystallization. In addition, it was also retrieved the laser power threshold which we observed experimentally. Finally, these findings prove the capability of Direct Laser Writing to pattern three-dimensional YIG structures with nanoscale resolution.

Regarding the minimum achievable patterning depth, the fine-tuning of the laser power around the threshold could potentially reduce it to a few nanometres, though this requires precise control due to the temperature gradient and possible variations in YIG's melting temperature near the surface. The maximum depth is on the other hand limited by the cracks and detachment formed at higher powers, suggesting that the upper limit is dictated by structural integrity. However, though optimized patterning strategies, such as reducing exposure time, could help extend this limit.

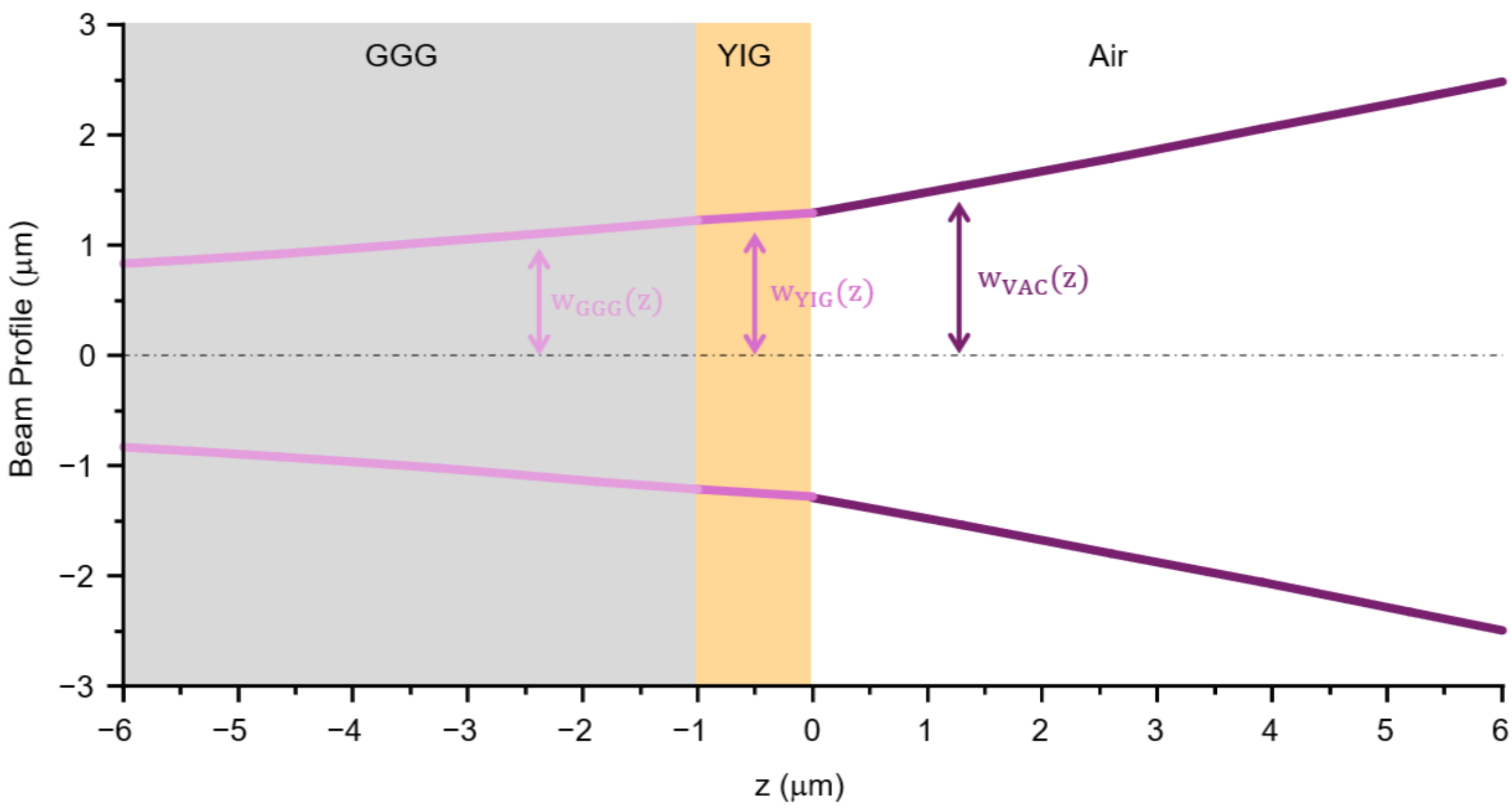

**Supplementary Figure 1: Laser beam profile through YIG/GGG.** Beam profile propagating through the system calculated as described in Supplementary Note 2. Deviations from the beam profile of a laser propagating only through air are due to the air/YIG and YIG/GGG interfaces, which present a change in the refractive index.

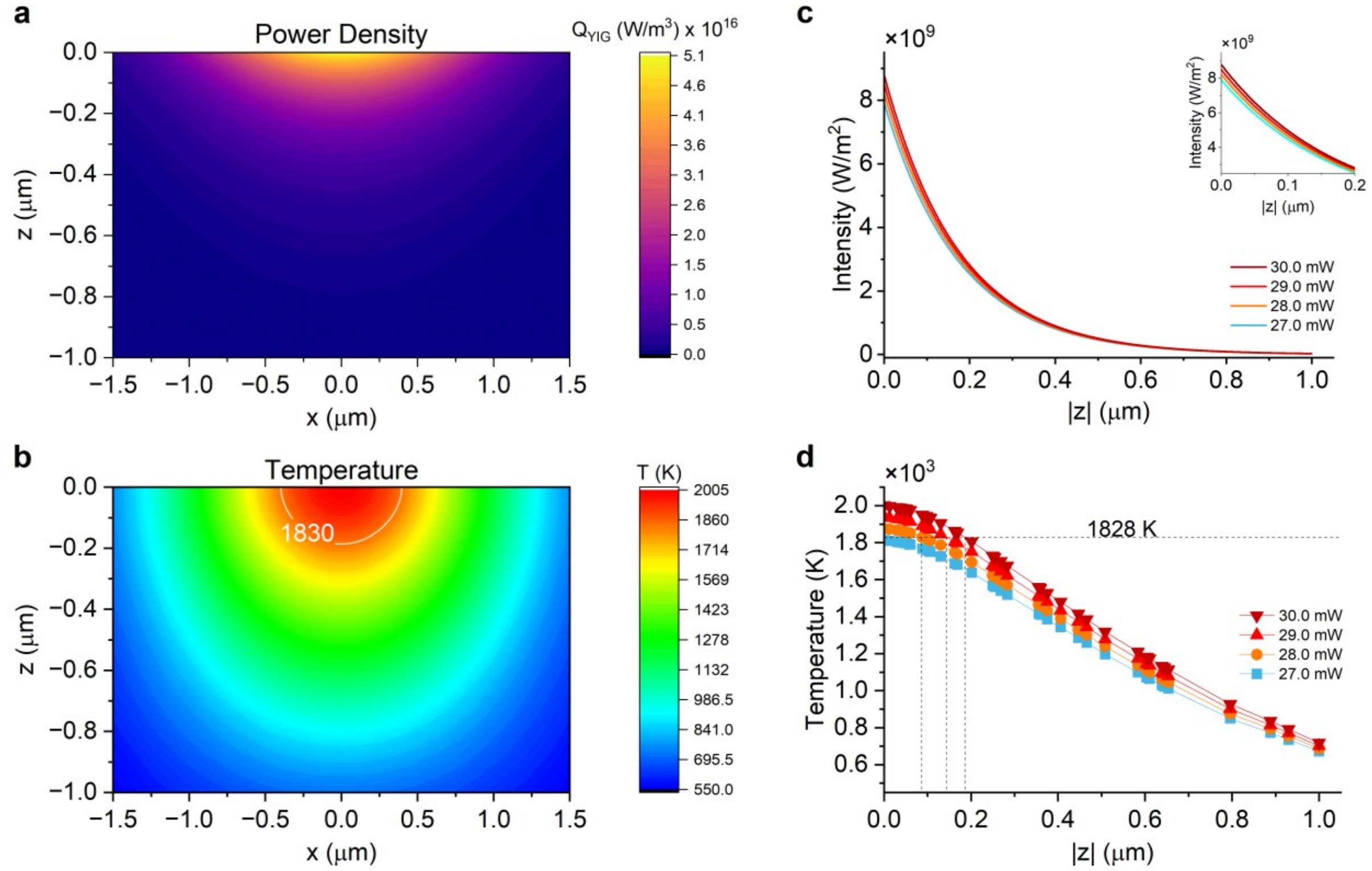

**Supplementary Figure 2: Simulated power and temperature gradients along the sample thickness. a,** Colormap of the power density in the *y* = 0 plane of the YIG layer, computed from equation (4) for a laser power $P_0$ = 30 mW. **b,** Optical intensity profiles along the thickness of the YIG layer, computed from equation (2), at *x* = *y* = 0, for different values or the laser power, from 27 mW to 30 mW. **c,** Colormap representing the temperature in the *y* = 0 plane of the YIG layer, for a laser power $P_0$ = 30 mW. Simulation time is 20 μs. The melting temperature of YIG is 1828 K. **d,** Temperature profiles along the thickness of the YIG layer, at *x* = *y* = 0, for different values or the laser power, from 27 mW to 30 mW. Simulation time is 20 μs. Extended details are discussed in Supplementary Note 2.

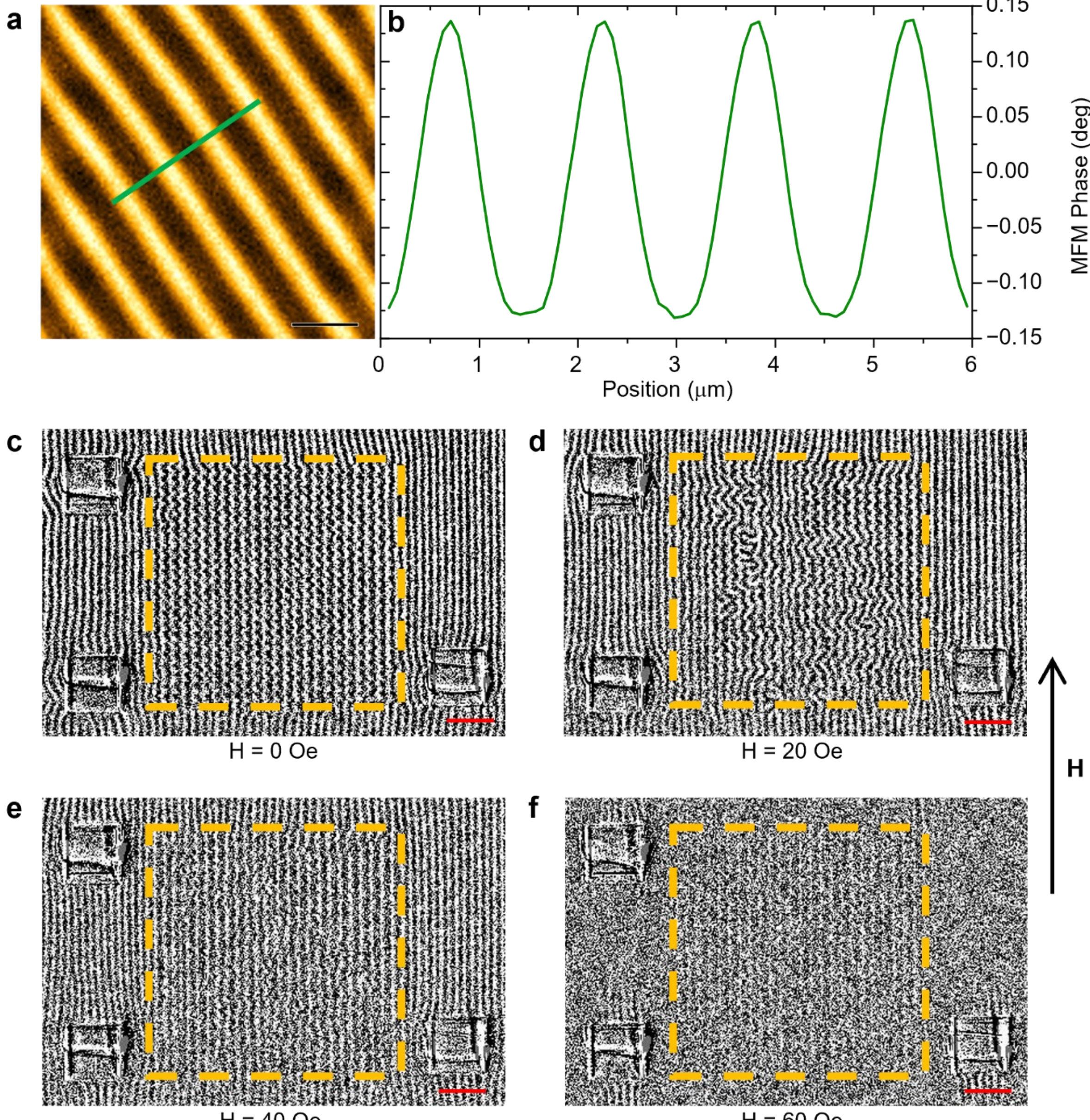

**Supplementary Figure 3: Zig-zag domains in array of nanodots**. **a,** Magnetic stripe domains measured by Magnetic Force Microscopy in the 1μm-thick pristine YIG film. Scale bar: 2 μm. **b,** MFM phase profile extracted along the green line in (**a**), revealing a periodicity of 1.6 μm. **c-f,** Array of nanodots with a lattice parameter of 2 μm (dashed squares) imaged by MOKE microscopy for different external fields, equal respectively to 0 mT (**c**), 2 mT (**d**), 4 mT (**e**), 6 mT (**f**). The periodic arrangement of irradiated points stabilize peculiar zig-zag domains at remanence, in contrast to straight stripe domains of the unpatterned film. By increasing the external field, the magnetic configuration further changes inhomogeneously before reaching saturation together with the pristine areas. A magnetic pinning effect can be seen near the square markers next to the dashed region. These observations prove the feasibility of manipulating the spin texture in YIG films by engineering the magnetic defects. In fact, a change in the magnetic properties of functional layouts lead eventually to a change in the magnetic configuration of neighbouring areas. Scale bars: 10 μm.

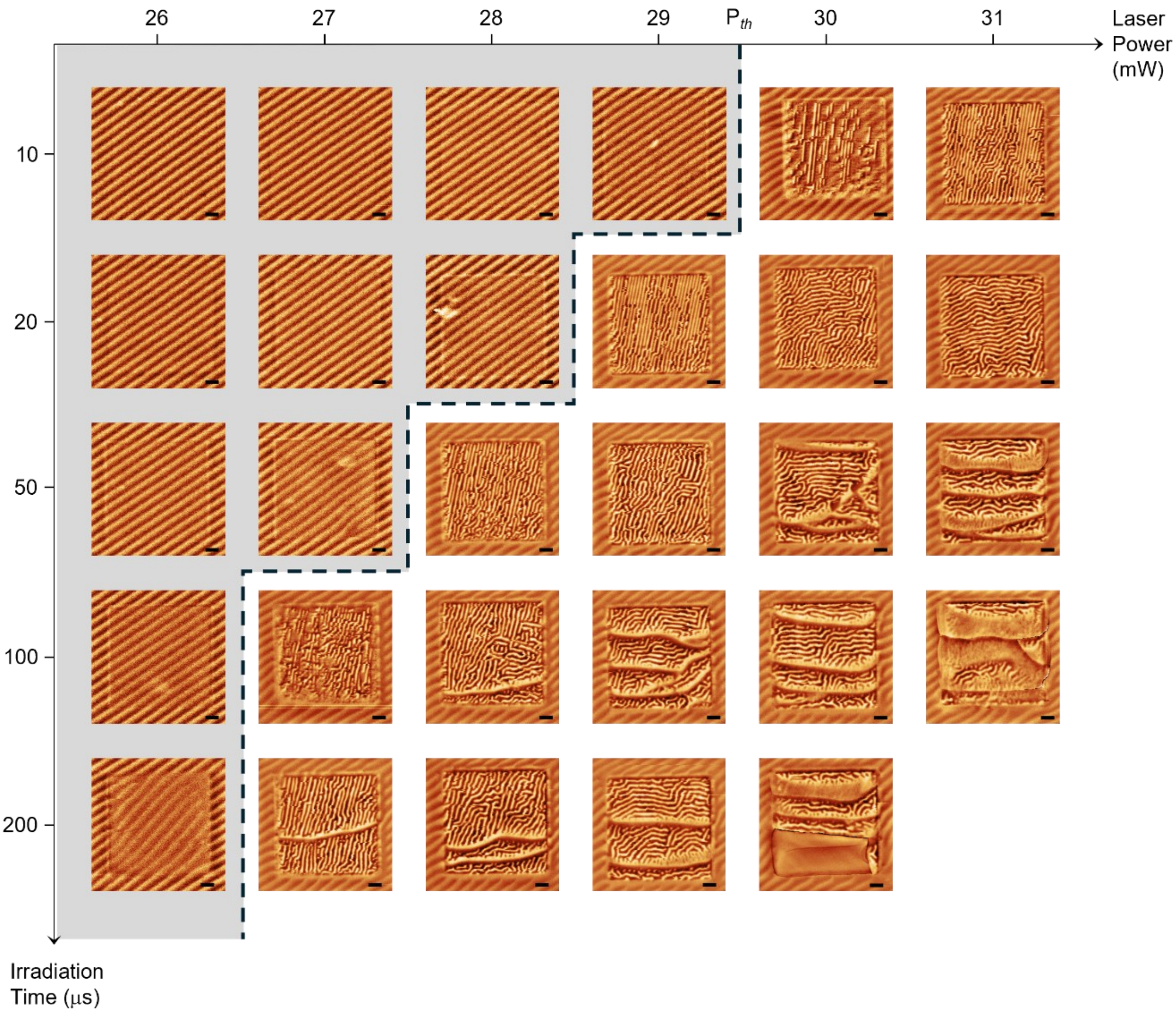

**Supplementary Figure 4**: **Effect of laser power vs irradiation time.** MFM images of 15um x 15um irradiated square patterns as a function of irradiation time and laser power. The power threshold changes with the exposure times and is represented by a dashed line. For exposure times longer than 20 μs, cracks are formed above threshold, which interrupt the stripe domains. Scale bars: 2μm.

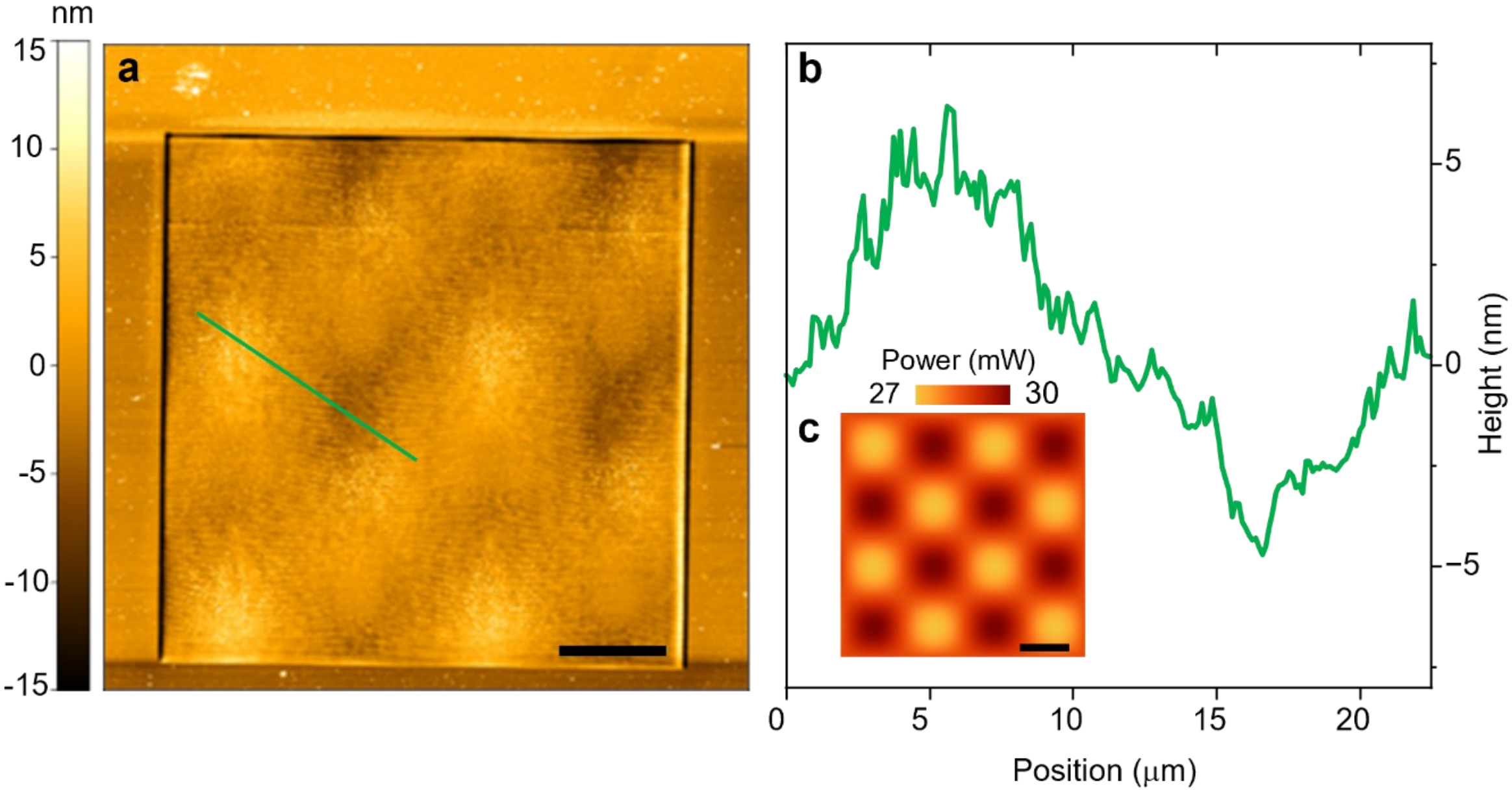

**Supplementary Figure 5: Topography maps of irradiated regions. a,** AFM image of the magnetic pattern in Figure 1f. The RMS roughness calculated outside and inside the square is equal respectively to 1.8 nm and 2.5 nm, indicating only a slight change in the overall surface quality. **b**, Topographic profile extracted along the green line in (**a**). **c**, Sinusoidal laser power map used for irradiating the pattern in (**a**). The maximum change in height within the patterned area is about 7 nm between the zone written at lower and higher power. The resulting morphology has approximately the same periodicity of the laser power map in (**c**). The dip of 15 nm at the edges of the irradiated regions is due to the non-idealities in the writing process and can be mitigated by adjusting the patterning parameters for pixels near the edges of the writing areas. Scale bars: 10 μm.

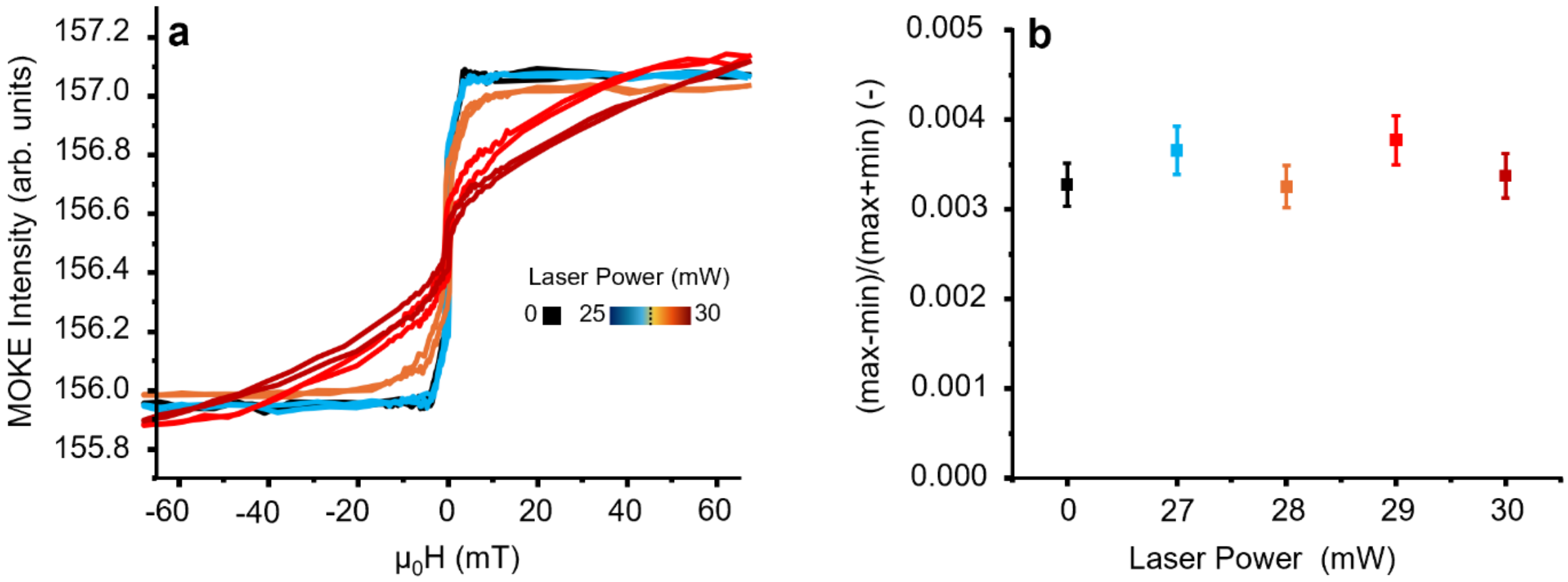

**Supplementary Figure 6: Non-normalized MOKE measurements**. **a,** Hysteresis loops presented in Figure 2a without data normalization. **b,** Normalized amplitude of each loop corresponding to different laser powers. The results show a negligible variability of the maximum MOKE intensity, suggesting that the saturation magnetization is affected negligibly by laser irradiation. Hence, the change in the magnetic configuration and hysteresis is primarily ascribed to a variation of the magnetic anisotropy. A graded change of $K_U$ is not compatible with the experimental SEM profile (Figure 3a) and with the thermally-activated threshold process described in Supplementary Note 2. Error bars in (**b**) are obtained as standard deviations of the signal beyond saturation at positive magnetic fields with a coverage factor equal to 5.

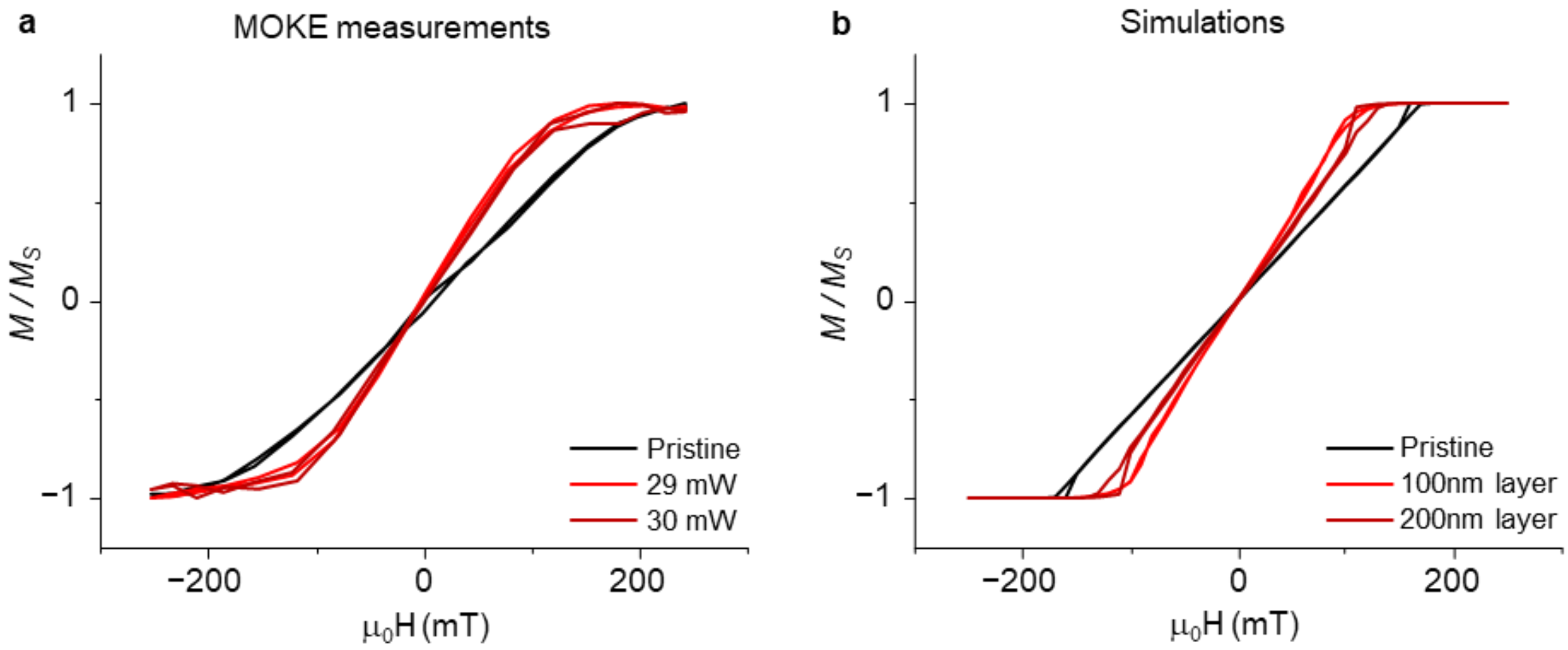

**Supplementary Figure 7: Out-of-plane MOKE measurements and micromagnetic simulations. a,** MOKE measurements performed with an out-of-plane magnetic field on pristine material (black curve) and two irradiated areas (red curves). **b,** Simulations of out-of-plane hysteresis loops for pristine film and irradiated systems consisting in a top modified layer with enhanced PMA and the pristine material underneath (see Supplementary Note 1 for the simulation procedure). The two red curves have been extracted from the output of micromagnetic simulations by averaging the magnetization over the cells in the top layer with the increased anisotropy. Indeed, the penetration depth of the 405 nm laser used for MOKE measurements is shallow enough to ensure that the signal originates solely from the top layer. Experimental data and simulations exhibit a good agreement.

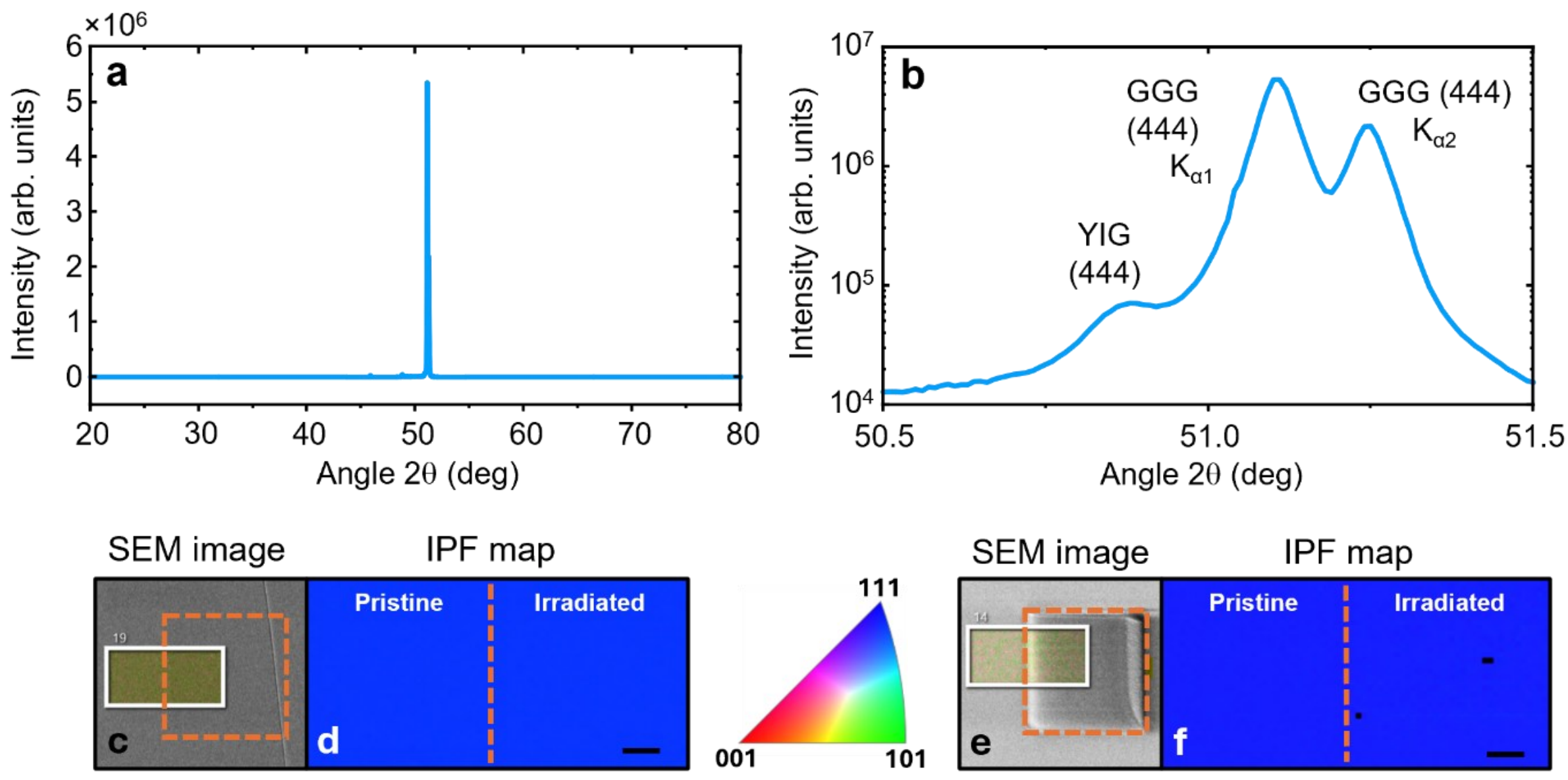

**Supplementary Figure 8: Crystallographic characterization of pristine and irradiated regions. a,** X-Ray Diffraction pattern of 1µm-thick YIG grown by LPE on GGG. **b,** Zoomed view in logarithmic scale of the maximum peaks in (**a**) with the identification of the characteristic (111) crystallographic orientation of the sample.[8] **c,d,** SEM image (**c**) and inverse pole figure map (IPF) obtained via Electron Backscattered Diffraction (EBSD) (**d**) acquired across the edge (white rectangle) of a patterned square (dashed region) irradiated with a laser power of 28 mW (just above threshold). **e,f,** SEM image (**e**) and EBSD IPF map (**f**) acquired across the edge (white rectangle) of a patterned square (dashed region) irradiated with a laser power of 31 mW (well above threshold, with some cracks appearing at the right side). Importantly, EBSD probing depth is between 10 and 50 nm, allowing to directly probe the modified top layer with no contribution from the bottom non-modified layer. In both cases, the crystallographic orientation inside the patterns is unchanged compared to the pristine film. This confirms the capability of direct Laser Writing to control the magnetic properties of the system without significantly altering its crystalline structure, which is a critical task in YIG nanopatterning. Scale bars: 1 µm.

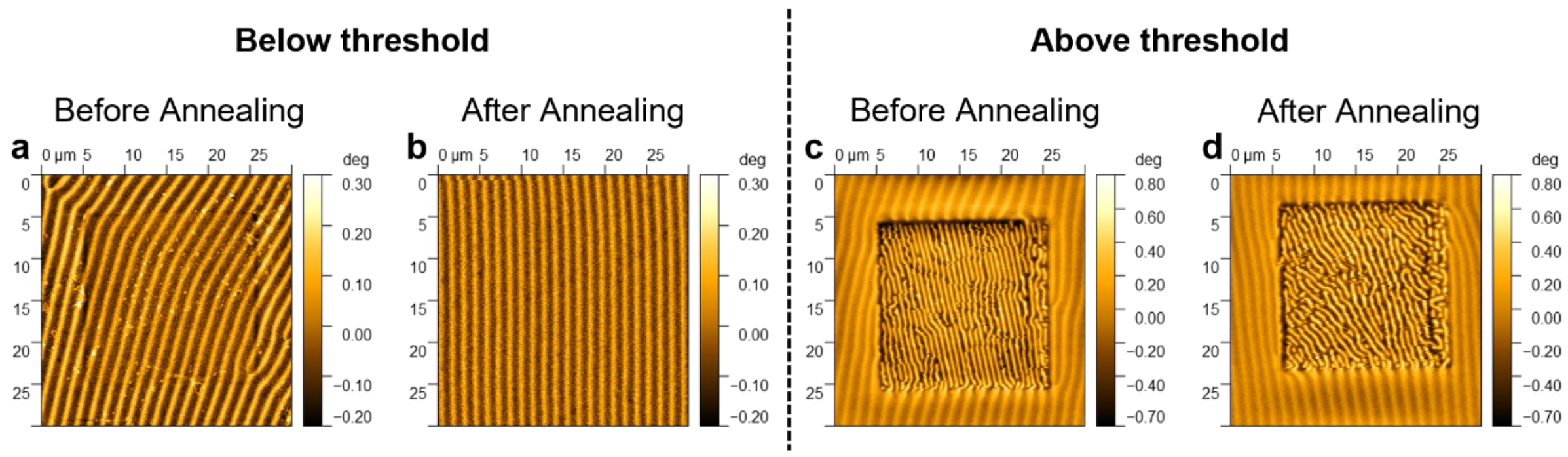

**Supplementary Figure 9: Effect of thermal annealing on irradiated regions.** MFM image of two squares patterned with laser power below threshold (**a,b**) and above threshold (**c,d**), before (**a,c**) and after (**b,d**) the annealing treatment, performed with a UniTemp RTP-150-HV, for 300 s at 850°C in 0.5 atm of oxygen. The squares written by laser irradiation below threshold are no longer distinguishable after the annealing, likely due to the filling or redistribution of the oxygen vacancies throughout the material. Conversely, the squares patterned above the threshold power remain clearly visible, proving the stability of the modification. This stability confirms the major role of structural strain for the enhancement of the PMA, along with the creation of oxygen vacancies. Notably, the size of the stripe domains within the pattern has increased after the annealing. This can be attributed to either partial strain relaxation or a reduction in the magnetostriction coefficient, resulting from the increased oxygen content following the thermal treatment, which in turn lower the anisotropy constant.

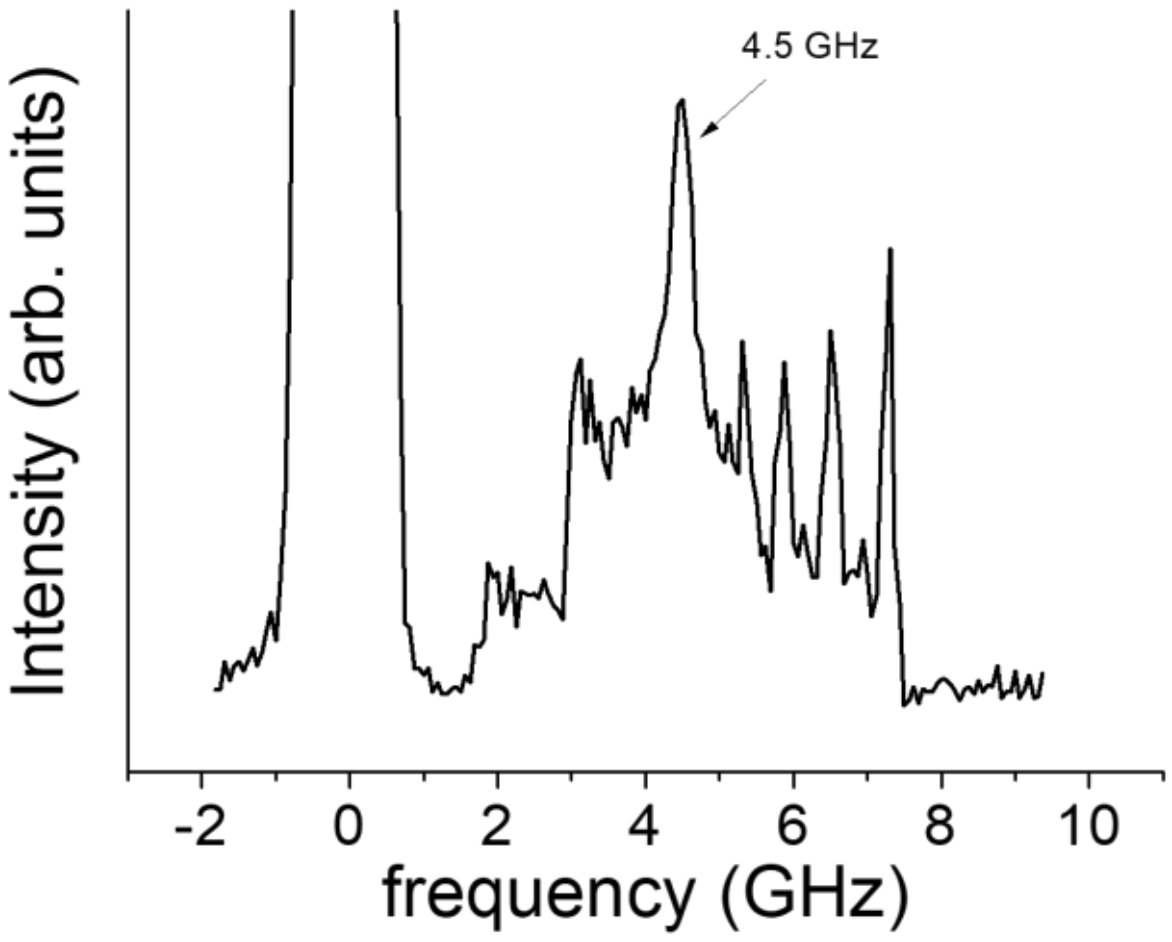

**Supplementary Figure 10: Micro-BLS measurement of thermal spin waves in pristine YIG film.**

The spectrum of thermal spin waves (naturally existing in ferromagnetic materials) under the effect of an in-plane magnetic field $\mu_0 H_{ext}$ =100 mT is acquired at normal incidence, measuring spin-wave modes around $k$=0. The intense peak measured at about 4.5 GHz corresponds to the FMR frequency, while the peaks in the range between 5 and 8 GHz correspond to perpendicular standing spin wave modes having 8-12 nodes along the film thickness.

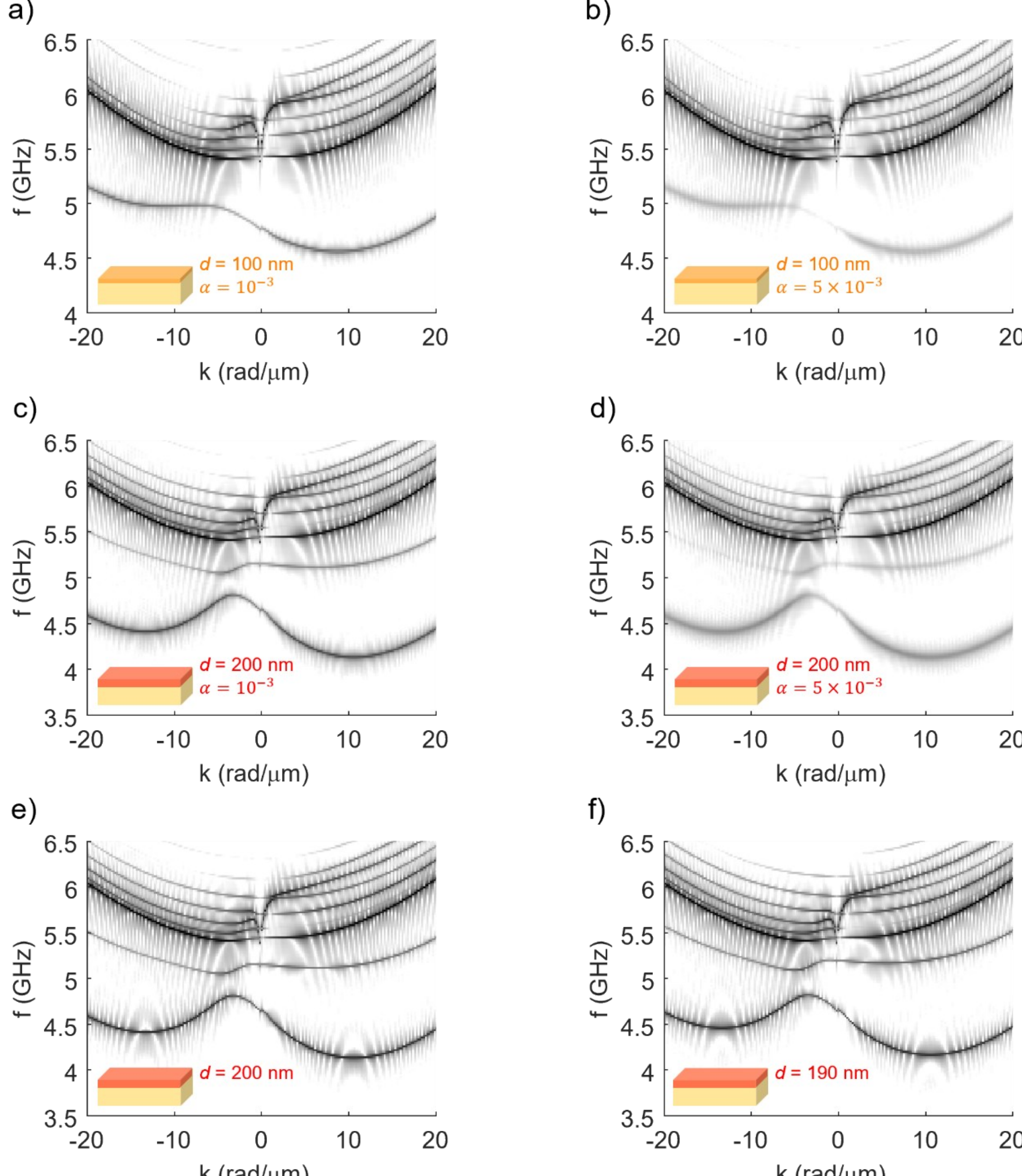

**Supplementary Figure 11: Simulated dispersion relations of bilayer systems with variable damping and thickness.** The top layer of the structure corresponds to YIG, which exhibits an increased anisotropy ($K_U$ = 6 kJ/m³) compared to the pristine bottom layer ($K_U$ = 0.2 kJ/m³, $\alpha$ = $10^{-4}$). **a,b,** Dispersion relations for a bilayer system with a 100 nm thick top layer having a damping value of $\alpha$ = $10^{-3}$ (**a**) and $\alpha$ = 5×$10^{-3}$(**b**). **c,d,** Dispersion relations for a bilayer system with a 200 nm thick top layer having a damping value of $\alpha$ = $10^{-3}$ (**c**) and α = 5×$10^{-3}$ (**d**). Panels **a-d** show that the effect of a larger damping in top layer is substantially different compared to the effect of a larger PMA. In fact, while an enhanced PMA along with the layer depth effectively tailor the shape of the bands, a change in α only entails a decrease in the intensity of the low-frequency modes, which are mostly localized in the top layer. **e,f,** Dispersion relations for two bilayer systems with top layer thickness of 200 nm (**e**) and 190 nm (**f**), while keeping a fixed 800 nm-thick pristine bottom layer. The results indicate that slight topographical variations in the top layer thickness have a negligible effect on the dispersion relation. The simulation procedure and micromagnetic parameters (unless otherwise specified) are detailed in Supplementary Note 1.